\documentclass[12pt,preprint]{aastex61}

\newcommand{\lapprox} {\, \lower3pt\hbox{$\sim$}\llap{\raise2pt\hbox{$<$}}\,}
\newcommand{\gapprox} {\, \lower3pt\hbox{$\sim$}\llap{\raise2pt\hbox{$>$}}\,}

\usepackage{amsmath}
\usepackage{natbib}
\usepackage{epstopdf}
\usepackage{float}
\usepackage[caption = false]{subfig}
\usepackage{graphicx}
\usepackage{listings}
\usepackage{amssymb}
\usepackage{amsmath,amsthm}
\usepackage{mathtools}
\usepackage{gensymb}
\usepackage{algorithm}
\usepackage{algorithmic}
\usepackage{multirow}
\usepackage{verbatim}
\usepackage{subfig}
\usepackage{xcolor}
\usepackage[bookmarks=false]{hyperref}

\begin{document}

\title{Identification of multiple hard X-ray sources in solar flares: \\ A Bayesian analysis of the February 20 2002 event}

\correspondingauthor{Michele Piana}
\email{piana@dima.unige.it}

\author{Federica Sciacchitano}
\affil{Dipartimento di Matematica, Universit\`a di Genova, via Dodecaneso 35 16146 Genova, Italy}

\author{Alberto Sorrentino}
\affil{Dipartimento di Matematica, Universit\`a di Genova, via Dodecaneso 35 16146 Genova, Italy and CNR - SPIN Genova, via Dodecaneso 33 16146 Genova, Italy}

\author{A Gordon Emslie}
\affil{Department of Physics and Astronomy, Western Kentucky University, Bowling Green, KY 42101, USA}

\author{Anna Maria Massone}
\affiliation{CNR - SPIN Genova, via Dodecaneso 33 16146 Genova, Italy}

\author{Michele Piana}
\affil{Dipartimento di Matematica Universit\`a di Genova via Dodecaneso 35 16146 Genova Italy and CNR - SPIN Genova, via Dodecaneso 33 16146 Genova, Italy}

\begin{abstract}
The hard X-ray emission in a solar flare is typically characterized by a number of discrete sources, each with its own spectral, temporal, and spatial variability.  Establishing the relationship amongst these sources is critical to determine the role of each in the energy release and transport processes that occur within the flare.  In this paper we present a novel method to identify and characterize each source of hard X-ray emission.  The method permits a quantitative determination of the most likely number of subsources present, and of the relative probabilities that the hard X-ray emission in a given subregion of the flare is represented by a complicated multiple source structure or by a simpler single source. We apply the method to a well-studied flare on 2002~February~20 in order to assess competing claims as to the number of chromospheric footpoint sources present, and hence to the complexity of the underlying magnetic geometry/toplogy.  Contrary to previous claims of the need for multiple sources to account for the chromospheric hard X-ray emission at different locations and times, we find that a simple two-footpoint-plus-coronal-source model is the most probable explanation for the data.  We also find that one of the footpoint sources moves quite rapidly throughout the event, a factor that presumably complicated previous analyses.  The inferred velocity of the footpoint corresponds to a very high induced electric field, compatible with those in thin reconnecting current sheets.
\end{abstract}


\section{Introduction} \label{sec:intro}

The \emph{RHESSI} instrument \citep{2002SoPh..210....3L} was designed to investigate particle acceleration and energy release in solar flares through imaging spectroscopy of hard X-ray and gamma-ray emissions from $\simeq 3$~keV to $\simeq 20$~MeV.  Over the course of its 15-plus-year lifetime, \emph{RHESSI} has provided thousands of hard X-ray images of solar flares. Modeling of the often complex structures thus revealed \citep{2003ApJ...595L.107E}, and relating them to theoretical models of the spatial distributions of hard X-rays in solar flares \citep[e.g.,][]{1971SoPh...18..489B,1973SoPh...31..143B,1975SoPh...41..135B,1987SoPh..107..263E} typically involves the identification of subregions of the image and the classification of each subregion into a particular ``type'' of source, such as extended coronal sources \citep[e.g.,][]{2007ApJ...669L..49K,2008A&ARv..16..155K,2008ApJ...673.1181K}, ``thick-target'' nonthermal sources confined to the corona \citep{2008ApJ...673..576X,2012A&A...543A..53G,2012ApJ...755...32G}  and chromospheric footpoints \citep[e.g.,][]{2002SoPh..210..373B,2002SoPh..210..383A,2004cosp...35.3787S,2008A&A...489L..57K}.

A particularly interesting example of the importance of such sub-source identification is the C7.5 flare observed by {\it RHESSI} on 2002 February 20, at $\sim$11:06 UT. This was one of the first solar flare events observed by \emph{RHESSI} and its characteristics, particularly its morphology, have been extensively studied, and we now summarize the results of these studies.

\begin{enumerate}

\item Figure~3 and 4 of \cite{2002SoPh..210..245S} show images of the event produced using the Clean algorithm \citep{2009ApJ...698.2131D} and the MEM-Sato maximum entropy method \citep{1999PASJ...51..127S}.  Both image reconstruction methods reveal the presence of two chromospheric footpoints (which become more prominent as the observing energy increases), in addition to a remote looptop source \citep{1995PASJ...47..677M}.

\item Figure~2 of \cite{2004SoPh..219..149A} shows further evidence for a two-footpoint structure, using five different imaging methods, including forward-fit \citep{2002SoPh..210..193A}, Clean \citep{2009ApJ...698.2131D}, Pixon \citep{1996ApJ...466..585M}, and two different maximum-entropy methods, one operating in $(x,y)$ space and the other in visibility (i.e., Fourier) $(u,v)$ space.

\item Figure~2 of \cite{2014ApJ...789...71F} again shows evidence for two chromospheric footpoints, a morphology also supported by SOHO EIT \citep{1995SoPh..162..291D} observations.

\item Figure~4 of \cite{2002SoPh..210..383A} shows the source centroids at different hard X-ray energies all concentrated in these two footpoints.

\item Adding to the complexity, Figure~1 of \cite{2005ApJ...621..482V} shows ``two distinct footpoints and a faint loop-top source'' located some 20 arcseconds toward the limb compared to the footpoints.

\item In their analysis of this event, \cite{2011ApJ...728....4G} discuss (cf. their Figure~3) ``a complicated pattern of low- and high-energy sources.''

\item \cite{2002SoPh..210..229K} discuss the possibility of a more complicated morphology with {\it three} footpoint sources, plus the looptop source.  Their Figure~3 shows that one of the footpoints (``Source 2'') disappears between 11:06:11 UT and 11:06:17 UT, and is promptly replaced by a ``Source 3'' located some 10~arcseconds to the South \citep[their Figure~4; see also Figure~2 of][]{2002SoPh..210..261V}. Both Source~2 and Source~3 are close to a location involving the disappearance of a small sunspot ``located near the edge of the dominant hard X-ray kernel'' \citep{2002ApJ...580L.177W}, providing evidence for energy release to due to magnetic reconnection/reconfiguration and possible evidence for an acceleration site in the chromosphere \citep{2009A&A...508..993B}.

\item \cite{2007ApJ...665..846P} show images of the event based both on observed counts (their Figure~1); they also performed a spectral inversion of the observed count visibilities to obtain ``electron visibilities'' and so produced images of the source in the electron domain (their Figure~5).  Because of the regularization in the spectral inversion step, the electron images vary much more smoothly with energy than the original count-based images.  It is therefore especially interesting that while the count-based images in \cite{2007ApJ...665..846P} do suggest the possibility of a third footpoint source (see especially the 34-38~keV image in their Figure~1), there is little evidence for this additional footpoint in the electron images.

\end{enumerate}

In light of these various results, assessing the quantitative likelihood (and significance) of a third footpoint source is an intriguing problem.  Further, \cite{2003ApJ...595L.103K} have shown that one of the footpoints in another event on 2003~July~23 moves across the solar disk with a speed that correlates well with the hard X-ray intensity of that footpoint source, suggesting that the rate of electron acceleration is correlated with the speed of the footpoint motion.  Indeed, one can easily hypothesize that higher footpoint speeds result in higher induced electric fields $\mathbf {E} = (\mathbf{v}/c) \times \mathbf{B}$ and hence greater acceleration rates.
There is therefore a clear need for a robust algorithm which can, amongst other things, distinguish between the alternative hypotheses of (a) the disappearance of one source and appearance of another, and (b) the drift of a single footpoint source. The development of just such an algorithm is the focus of the present work.

\emph{RHESSI} uses indirect, rotating-collimator-based imaging techniques \citep{2002SoPh..210...61H}. The ``native'' output of the imaging data is in the form of count modulation profiles from each of \emph{RHESSI}'s nine detectors, which can be temporally stacked to produce a set of ``visibilities'' (two-dimensional spatial Fourier components of the image), with each detector sampling Fourier components tracing a circle in $(u,v)$ space over the course of the $\simeq$4~s spacecraft rotation period.  Reconstruction of images then proceeds in one of several ways:

\begin{itemize}
\item Fourier transformation (``backprojection'') of the data to produce a ``dirty'' image in $(x,y)$ space, which is then improved using methods like Maximum Entropy \citep{2006ApJ...636.1159B}, Clean \citep{1974A&AS...15..417H,2009ApJ...698.2131D}, or Pixon \citep{1996ApJ...466..585M};
\item Inversion of count modulation profiles by means of optimally stopped iterative schemes \citep{2013A&A...555A..61B};
\item Interpolating the (relatively sparse) visibility data to obtain a smooth surface in $(u,v)$ space, which can then be Fourier-transformed to produce an image in the $(x,y)$ domain \citep{2009ApJ...703.2004M};
\item Reconstruction methods from visibilities using compressed sensing approaches \citep{duval2017compressed,2017arXiv170803877D,2017ApJ...849...10F}
\item Forward-fitting the visibilities with a parametric model in order to obtain the best-fit parameters \citep{2002SoPh..210..193A}.

\end{itemize}

The first four items essentially refer to \textit{image reconstruction} methods that involve the reconstruction from data of an entire image, i.e., a set of pixel values that correspond to the intensity of the hard X-ray emission, at the photon energy of observation, in an area element on the solar disk.  By contrast, the last method represents an alternative approach, i.e. it is a \textit{parameter estimation} method, where it is assumed that the data (again, count modulation profiles or visibilities) have been generated by a very small number of simply-shaped geometrical objects. The parameters of these objects are then estimated through non-linear optimization techniques. To the best of our knowledge, to date this visibility forward fitting (VFF) algorithm \citep{2008ApJ...673..576X} is the only algorithm that provides statistical uncertainties of fitted source parameters; however, the VFF method is limited by the need for {\it a priori} knowledge of the number and the shapes of sources.  Furthermore, in the current default version of VFF provided in the SSW software, only a single circular or elliptical Gaussian, or single loop, or two circular Gaussians can be reconstructed, although more recently the VVF algorithm has been modified \citep{2009ApJ...698.2131D} such that it can address two elliptical sources simultaneously.

We here present a Bayesian subsource identification and characterization method that is ideally suited to addressing the issue of determining the number, and structure, of various sub-regions in a hard X-ray image.  The method quantitatively assesses the likely number of discrete sources in an image and the parameters that characterize each source, together with their quantitative uncertainties. This approach works with an arbitrary number of sources, the shapes of which are automatically determined by the method, together with the source parameters and their uncertainties. Importantly, the method also provides a quantitative way of determining the relative probability of configurations with different numbers of sources. It is therefore particularly useful in assessing the complexity (or lack thereof) in the source image, such as the number of footpoint sources in the 2002~February~20 event.

In Section~\ref{sec:Bayes} we describe the method.  In Section~\ref{sec:20_feb} we describe the \emph{RHESSI} observations used for the analysis, i.e. the relatively well-studied 2002 February 20 11:06 UT event, and we compare the obtained results with the ones described in \citep{2002SoPh..210..245S,2002SoPh..210..229K}. Section \ref{sec:impact} discusses the impact of the prior distribution on the results.  Our conclusions are offered in Section~\ref{sec:disc_concl}.


\section{Bayesian parametric estimation of solar flare structures}\label{sec:Bayes}

In this section we describe the source parametrization, the Bayesian model and the Monte Carlo algorithm that we use to estimate the flare parameters from the visibilities measured by \emph{RHESSI} (this same approach can be extended to the Bayesian fitting of \emph{RHESSI} count modulation profiles).

\subsection{Parametrization of flare sources}

Hard X-ray images in solar flares \citep[e.g.,][]{2003ApJ...595L.107E,2008ApJ...673..576X} can involve either a single source (compact or extended), compact sources, extended sources, or combinations thereof. Very often, such maps can be effectively represented as the superposition of different two-dimensional distributions, which we assume to be each described by a simple geometric object, such as a circle, an ellipse, or a ``loop''.  Even though the parametrization we shall use to describe these objects is not entirely new \citep{2002SoPh..210..193A}, for the sake of completeness we provide here a brief description of it.

\begin{figure}[t]
\hspace{-3.0cm}
\begin{center}
\includegraphics{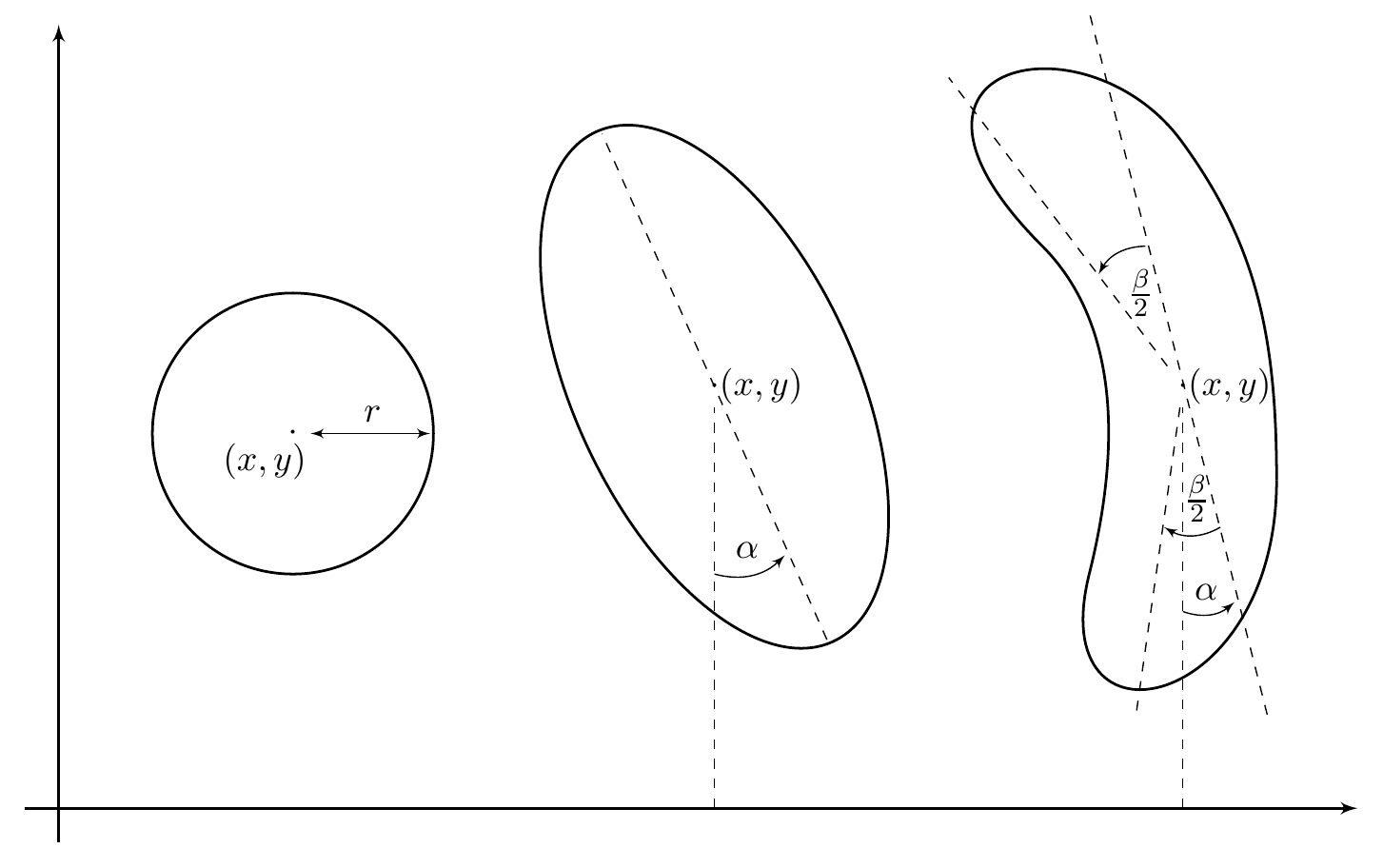}
\end{center}
\caption{The three source types: circle, ellipse and loop }
\label{fig:param}
\end{figure}

The various model sources used are characterized as follows (see Figure~\ref{fig:param}):

\begin{itemize}

\item {\it Circle:} a circular source is defined by the $x$ and $y$ positions of its center, measured in arcseconds from the center of the image, the flux $\phi>0$ (which represents the integrated intensity over the source), and the full width at half maximum (FWHM) intensity $2r>0$ (where $r$ measures the radius of the source).

\item {\it Ellipse:} an ellipse is defined by the location $(x,y)$ of its center, the semi-major axis $a$ corresponding to the 50\% intensity level that extends furthest from the center, the rotation angle $\alpha \in [0,180\degree]$ of that direction, and the eccentricity

$$\varepsilon= \sqrt{1- \frac{b^2}{a^2}} \,\,\, ,$$

where $a>0$ and $b>0$ correspond to the length of the semi-major and semi-minor axes, respectively.

\item {\it Loop:} a ``loop'' is defined by the above ellipse parameters plus an additional parameter $\beta \in [-180\degree,180\degree]$ which indicates the degree of curvature of the loop shape (see Figure~\ref{fig:param}). The angle $\beta$ is inversely proportional to the radius of curvature; we choose to use the reciprocal curvature parameter $\beta$ to avoid the problem of dealing with infinite values of curvature for the elliptical and circular sources.

\end{itemize}

All these sources are thus defined by the following vector

\begin{equation}
\label{eq:par_theta}
\theta= (x,y, \phi, r, \alpha, \varepsilon, \beta),
\end{equation}
where the last three parameters can possibly (for the case of a circle or an ellipse) be zero or indeterminate.  Figure \ref{fig:param} provides a visual description of the parameters.

The parametrization in Equation~\eqref{eq:par_theta} has already been presented in \cite{2002SoPh..210..193A}, where a forward-fitting method to reconstruct hard X-ray images of the solar flares was described. In their study, the authors recovered the best estimates of the various source parameters together with error bars on the estimated parameters. However, a major limitation of the \cite{2002SoPh..210..193A} method is that the number and shape of the sources have to be assumed {\it a priori}. Furthermore, the algorithm is able to recover only the following limited configurations: at the time of writing, a single circle, a single ellipse, a single loop, or two circles. Here we overcome such limitations, by explicitly embodying the number $N$ of sources and the different source types among the unknowns to be determined.  Specifically, in the approach to be described the goal is to probabilistically infer the following object:

\begin{equation}
	X=(N,T_{1:N}, \theta_{1:N})
	\label{source_model}
\end{equation}
where $N=0, \dots, N_{\max}$ represents the number of sources, $T_{1:N}=(T_1, \dots, T_N)$ the source types and $\theta_{1:N}=(\theta_1, \dots, \theta_N)$ the source parameters for each source.

\subsection{Bayesian model}

In a Bayesian setting, all the variables of the problem are considered as random variables, with information on their values encoded as their probability distributions. In the following we will for simplicity use lower-case letters to indicate both the random variables and their specific values in a given instance. We are interested in the posterior distribution for flare source parameters, as described by Equation~\eqref{source_model}, conditioned on the data, namely a set of observed visibilities $y$.  From Bayes' theorem we have

\begin{equation}
 \label{bay}
 \pi_{\text{po}}(x|y)=\frac{\mathcal{L}(y|x)\pi_{\text{pr}}(x)}{p(y)},
\end{equation}
where  $\pi_{\text{po}}(x|y)$ is the posterior probability distribution representing the conditional probability density of $x$ given the data $y$, $\mathcal{L}(y|x)$ is the likelihood function that expresses the likelihood that the data $y$ is indeed due to the image $x$, $\pi_{\text{pr}}(x)$ is the prior probability density which describes the properties of the source configuration that we want to recover, and $p(y)$ is a normalization factor, independent of $x$. In the following we will describe our choice of the prior density and the algorithm we use to then approximate the posterior density. We notice that the normalization factor $p(y)$ is generally unknown, but its knowledge is not necessary to apply most Monte Carlo algorithms and, specifically, the Monte Carlo algorithm described below.

Since the number of sources is not known {\it a priori}, we need to consider a variable-dimension model, i.e., the state space $\chi$ of the sources is defined as follows:

\begin{equation}
\label{eq:space}
\chi \coloneqq \displaystyle\cup_{k=1}^N \{k\}\times \Theta_k \,\,\, ,
\end{equation}
where $\Theta_k$ is the state space for source number $k$ of the $N$ sources present.

\subsubsection{The prior distribution}\label{subsec:prior}

As in all Bayesian problems, the choice of the prior distribution $\pi_{\text{pr}}(x)$, which is the starting point for the calculation of subsequent improved distributions using the available data, is critical.  Our choice of suitable prior probability densities $\pi_{\text{pr}}(x)$ reflects basic expectations that are based on available knowledge of solar flare geometries. By assuming independence between the number of sources, the source types and the source parameter values, the prior distribution $\pi_{\text{pr}}$ can be separated as follows:

\begin{equation}\label{eq:prior}
\pi_{\text{pr}}(X)= \pi_{\text{pr}}(N, T_{1:N}, \theta_{1:N})= \rho_1(N) \, \prod_{k=1}^N \rho_2(T_k) \, \rho_3(\theta_k) \,\,\, ,
\end{equation}
where the $\rho_i$ ($i \in \{ 1,2,3 \}$) indicate the prior distributions of the number of sources, the source types, and the source parameters, respectively. We remark that the assumption of independence of these quantities is not simply useful; rather, it corresponds to the minimization of the information content of the prior.

We now further assume that the number of sources in a map is Poisson distributed with mean $\lambda>0$:

\begin{equation}\label{eq:prior-1}
\rho_1(N)=\frac{e^{-\lambda}\lambda^N}{N!} \,\,\, ,
\end{equation}
and we will limit $\lambda \in [0,3]$. In addition, we shall limit $N$, the maximum number of sources, to $N \leq 5$, because past observations \citep[e.g.,][]{2003ApJ...595L.107E} suggest that, given the spatial resolution of \emph{RHESSI}, a higher number of distinct sources is unlikely.

For the source type we adopt the following prior distribution

\begin{equation}
\label{eq:pr_sh}
\rho_2(T_k)=
\begin{cases}
p_C \quad &\text{if } T_k \text{ is a circle} \\
p_E \quad &\text{if } T_k \text{ is an ellipse} \\
p_L \quad &\text{if } T_k \text{ is a loop},
\end{cases}
\end{equation}
where $p_C,p_E,p_L \in [0,1]$.  Now, in general the morphology of solar flare hard X-ray sources is dependent on the photon energy observed: images at lower energies $\lapprox 15$~keV generally are dominated by extended loop-like (or elliptical) structures, while images at higher photon energies are more likely to include compact sources (possibly exclusively so).  The former are typically interpreted as extended thermal sources \citep[see, however,][]{2008ApJ...673..576X,2012A&A...543A..53G,2012ApJ...755...32G}, while the latter are generally considered as chromospheric footpoint sources, which are compact because of the much higher density and so electron stopping distance \citep{1972SoPh...26..441B,1978ApJ...224..241E}.  In our analysis we will use $(p_C,p_E,p_L)=(\frac{1}{2}, \frac{1}{4}, \frac{1}{4})$ if the energy is high and $(p_C,p_E,p_L)=(\frac{1}{4}, \frac{1}{4}, \frac{1}{2})$ otherwise.  We stress that these are only assumed {\it prior} distributions; they do not represent the final (data-informed) result.

The individual source parameters are {\it a priori} assumed to be uniformly distributed in their respective ranges:

\begin{equation}
\label{eq:pr_par}
\begin{aligned}
(x,y) &\sim \mathcal{U} \left ( \left [ x_c-\frac{F}{2},x_c+\frac{F}{2} \right ], \left [ y_c-\frac{F}{2},y_c+\frac{F}{2} \right ] \right ) \\
r &\sim \mathcal{U} \left ( \left [ 0,\frac{F}{3} \right ] \right ) \\
\phi &\sim \mathcal{U} \left ( \left [ 0,\max Re(V) \right ] \right ) \\
\varepsilon & \sim \mathcal{U} \left ( \left [ 0,1 \right ] \right ) \\
\alpha &\sim \mathcal{U} \left ( \left [ 0,180\degree \right ] \right ) \\
\beta &\sim \mathcal{U} \left ( \left [ -180 \degree,180\degree \right ] \right ) \,\,\, ,
\end{aligned}
\end{equation}
where $(x_c,y_c)$ is the map center, $F$ indicates the field of view of the map in each direction and $V$ represents the (complex) visibilities.  Combining Equations~\eqref{eq:prior-1}, \eqref{eq:pr_sh}, and \eqref{eq:pr_par}, the prior distribution $\pi_{\text{pr}}$ is given by

$$\begin{aligned} \pi_{\text{pr}}&= \rho_1(N) \Biggl( \prod_{k=1}^N \rho_2(T_k)\Biggr)\Biggl( \prod_{k=1}^N \rho_3(\theta_k)\Biggr)\\
&= \frac{e^{-\lambda}\lambda^N}{N!} p_C^{N_C} p_E^{N_E} p_L^{N_L}  \biggl( \frac{1}{F}\biggr)^{2N}  \biggl( \frac{3}{F}\biggr)^{N} \\
& \qquad \qquad\biggl( \frac{1}{\max Re(V)}\biggr)^{N} \biggl( \frac{1}{180}\biggr)^{N_E+N_L}\biggl( \frac{1}{360}\biggr)^{N_L}.
\end{aligned}$$

\subsubsection{Likelihood function}
Since visibilities are linear combinations of independent photon count numbers \citep{2002SoPh..210...61H}, we can make the standard approximation that noise on the visibilities is Gaussian additive; the likelihood function is thus given by
\begin{equation}\label{likelihood}
\mathcal{L}(y|x) \propto \exp  \left(- \sum_i \frac{|y-A(x)|_i^2}{2 \sigma_i^2} \right)
\end{equation}
where $A$ indicates the forward operator, i.e. the projector from the imaging parameter space onto the visibility space.  The standard deviations $\sigma_i^2$ are assumed to be known; in the data analysis below, we will be using the uncertainty values associated to the measured visibilities as provided by the SolarSoftWare (SSW) IDL routines.

\subsection{Approximation of the posterior distribution}

The posterior probability density is a complicated function on a relatively high--dimensional space, so that dealing with it analytically is impractical.  We therefore use a Sequential Monte Carlo (SMC) algorithm that produces a sample set which is approximately distributed according to the posterior, and can thus be used to make inference on the values of the various parameters. A brief description of the SMC algorithm is given in the Appendix below; for more details, we refer to \cite{del2006sequential} for the original article in which the SMC method is introduced, and to \cite{sorrentino2014bayesian} for an example of its application to a mathematically similar inverse problem. 

Once the algorithm converges, the parameters of the reconstructed map are computed by using the following point estimates. For the number of sources $\hat{N}_i$ we use the most probable value, given by the MAP (Maximum a Posteriori) estimate. Similarly, the estimated locations of the sources are given by the local modes of the posterior distribution for the source locations, conditioned on the estimated number of sources; $\phi, r,$ and $\varepsilon$ are estimated by the mean values of the conditional distribution conditioned on the source location and number of parameters; the angles $\alpha, \beta$ are estimated as the modes of the corresponding conditional distributions. Finally, the types of the sources are estimated {\it a posteriori} by using the estimates of the parameters. In particular, for our experiments we use the following classification

\begin{equation}\label{eq:class}
\begin{cases}
 \text{circle;} &\text{ if }\varepsilon\leq 0.1 \\
 \text{ellipse;} &\text{ if } 0.1 \leq \varepsilon < 0.4, |\beta| \leq 45\degree\\
 &  \text{ and } \varepsilon \geq 0.4  \text{ and } |\beta| \leq 10\degree \\
  \text{loop;} & \text{otherwise.}
\end{cases}
\end{equation}

\section{Application to RHESSI data - the 2002 February 20 event}\label{sec:20_feb}

\subsection{General}

We now apply our analysis to the C7.5 flare observed by {\it RHESSI} on 2002 February 20, at $\sim$11:06 UT. This was one of the first solar flare events observed by \emph{RHESSI} and, as discussed in Section~\ref{sec:intro}, its characteristics, particularly its morphology, have been extensively studied, particularly with regard to the number, and nature, of different sources present as time progresses.

Our Bayesian subsource identification and characterization method is ideally suited to addressing the issue of the number of discrete subsources in the region. Using the method as described above, we can quantitatively assess both the number of subsources in an image and the parameters that characterize each source, together with their uncertainties. To the best of our knowledge, to date only the visibility forward fitting (VFF) algorithm \citep{2008ApJ...673..576X} is able to provide statistical uncertainties of fitted source parameters; however, as we have already pointed out, the VFF method is limited by the need for {\it a priori} knowledge of the number and the shapes of sources. Furthermore, in the current default version of VFF provided in the SSW software, only a single circular or elliptical Gaussian, or single loop, or two circular Gaussians can be reconstructed, although more recently the VVF algorithm has been modified \citep{2009ApJ...698.2131D} such that it can address two elliptical sources simultaneously.

In this section, we will compare the values obtained through both the new Bayesian method and the VFF. As we will see, the results are very similar; however the Bayesian method has two very important advantages: it does not require any {\it a priori} knowledge of the shapes and number of sources. Furthermore, it can recover every plausible configuration with the corresponding probability, and not only the simplest one as in the VFF algorithm.

In order to be consistent with the results in \cite{2002SoPh..210..245S} and \cite{2002SoPh..210..229K}, in our experiments, we use \emph{RHESSI} detectors 3 through 9, a field of view $F=64$~arcseconds; further, we choose map-centers at $[918,270]$ and $[912,263]$ for Sections~\ref{sec:sui} and~\ref{sec:kru}, which compare our results with those of the respective previous authors.  The number of particles is set to $N=10000$, the maximum number of sources admitted in a map is  $N_{\max}=5$ and, coherently to what done in VFF, the noise level is given by adding a systematic error  $\sigma_{\text{sys}}=0.05$ to the error $e$ computed by the \emph{RHESSI} SSW routines, i.e.,

$$ \sigma_{\text{noise}} = \sqrt{e^2  + \sigma_{\text{sys}}^2  V_{\text{amp}}^2} \,\,\, ,$$
where $V_{\text{amp}}$ is the absolute value of the visibilities. Therefore, the only parameter that needs to be tuned is $\lambda$, and we assumed values in the range $[1,3]$.  In most cases slightly changing $\lambda$ will not influence the reconstruction, but only the convergence speed; 
for more details, we refer the reader to Section \ref{sec:impact}.

In order to calculate the visibilities generated by a given source, which is needed to compute the likelihood of the Monte Carlo samples, we employ the SSW function \texttt{hsi\textunderscore vis\textunderscore fwdfit\textunderscore func}.

\subsection{Comparison with the results in Sui et al}\label{sec:sui}

In this section, we focus on the time interval 11:06:10-11:06:24~UT and several energy intervals spanning the range from 6~keV to 70~keV. We compare the results found in \cite{2002SoPh..210..245S} with those using our Bayesian approach.

In \cite{2002SoPh..210..245S}, the Clean  \citep[see][]{2002SoPh..210...61H} and the maximum entropy \citep[see][]{1999PASJ...51..127S} reconstruction techniques have been used to study the number of sources with different energy bands at a fixed time interval.  In particular, \cite{2002SoPh..210..245S} found (their Figures~3 and ~4) that the map in the low-energy band (6-10~keV) shows a single structure which is located between the two footpoints observed at higher energies. In the interval 10-14~keV, a thermal source and a possible loop source are present. In the energy bands between 14 and 50~keV there are two footpoints and the loop-top source, while in the last energy band 50-70~keV only the two footpoints are present and the loop-source is not visible.

\begin{figure*}[!h]
\begin{center}
\hspace{-1cm}\includegraphics[height=8.5cm]{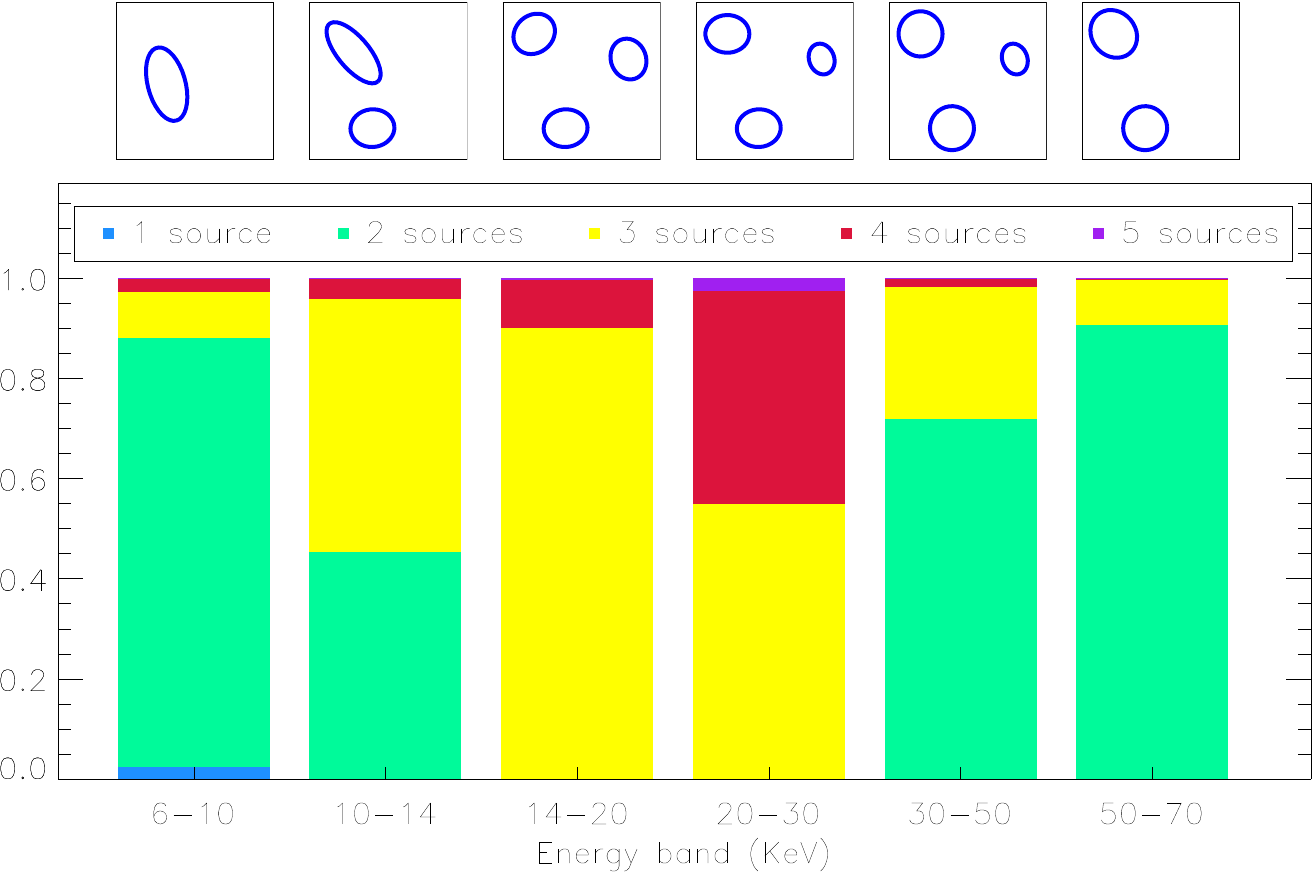}
\caption{Posterior probability for the number of sources, color-coded: in blue the probability of a single source, in green that of two sources, in yellow that of three sources, and so on. The parameter $\lambda$ has been set as follows: $\lambda=1$ for the energy band 6-10~keV, $\lambda=2$ for 10-14~keV and 50-70~keV, and $\lambda=3$ for 14-50~keV.}\label{fig:sui_num_sources}
\end{center}
\end{figure*}

For the prior distribution used in our approach (see equation (\ref{eq:prior})), we set the number of sources $\lambda$ according to the results found in \cite{2002SoPh..210..245S}, i.e., $\lambda=1$ for the energy band 6-10~keV,  $\lambda=2$ for 10-14~keV and 50-70~keV, and $\lambda=3$ for 14-50~keV.


\begin{figure*}[!h]
\begin{center}
\subfloat[6-10keV]{\includegraphics[height=4.5cm]{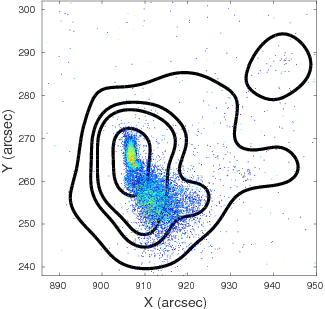}}
\subfloat[10-14keV]{\includegraphics[height=4.5cm]{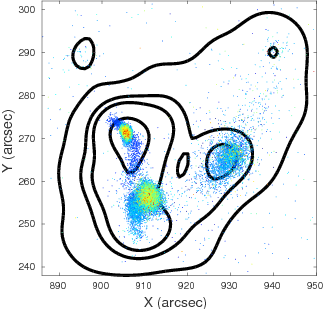}}
\subfloat[14-20keV]{\includegraphics[height=4.5cm]{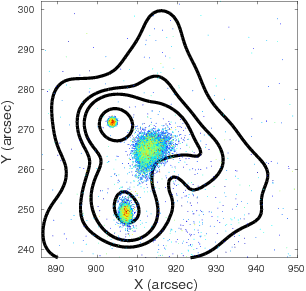}}\\
\subfloat[20-30keV]{\includegraphics[height=4.5cm]{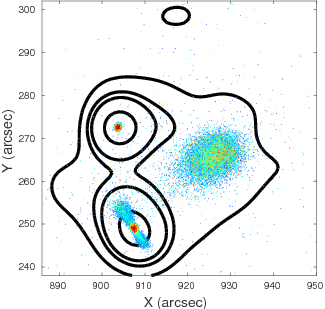}}
\subfloat[30-50keV]{\includegraphics[height=4.5cm]{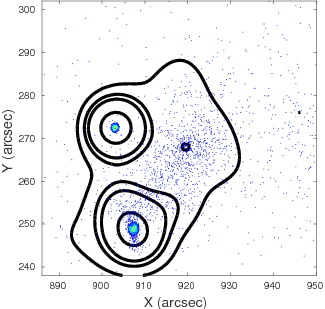}}
\subfloat[50-70keV]{\includegraphics[height=4.5cm]{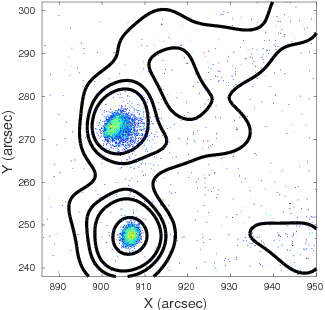}}\\
\end{center}
\caption{Distribution of the particles in the last iteration and (black lines) Clean reconstruction. The parameter $\lambda$ has been set as in Figure~\ref{fig:sui_num_sources}. Clean contour levels are 0.08, 0.2, 0.3, and 0.7. }
\label{fig:sui_xy}
\end{figure*}

\begin{figure*}[!h]
\begin{center}
\subfloat[6-10keV]{\includegraphics[height=4.5cm]{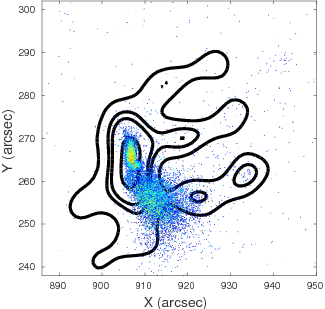}}
\subfloat[10-14keV]{\includegraphics[height=4.5cm]{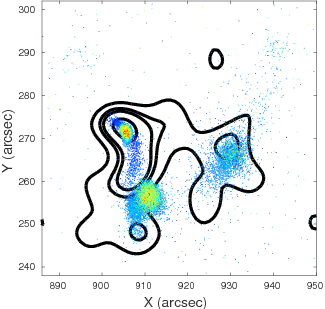}}
\subfloat[14-20keV]{\includegraphics[height=4.5cm]{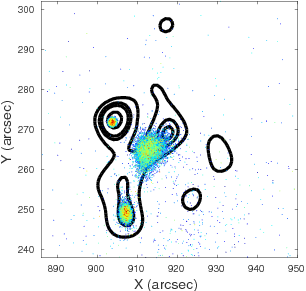}}\\
\subfloat[20-30keV]{\includegraphics[height=4.5cm]{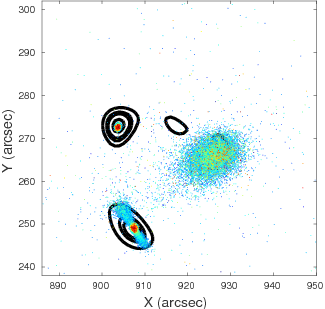}}
\subfloat[30-50keV]{\includegraphics[height=4.5cm]{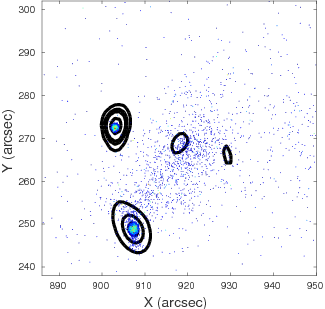}}
\subfloat[50-70keV]{\includegraphics[height=4.5cm]{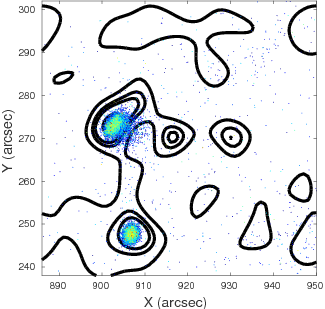}}\\
\end{center}
\caption{Distribution of the particles in the last iteration and (black lines) MEM-NJIT reconstruction. The parameter $\lambda$ has been set as in Figure~\ref{fig:sui_num_sources}. MEM-NJIT contour levels are 0.08, 0.2, 0.3, and 0.7. }\label{fig:sui_xy_mem}
\end{figure*}

\begin{figure*}[!h]
\begin{center}
\subfloat[Clean]{\includegraphics[height=4.5cm]{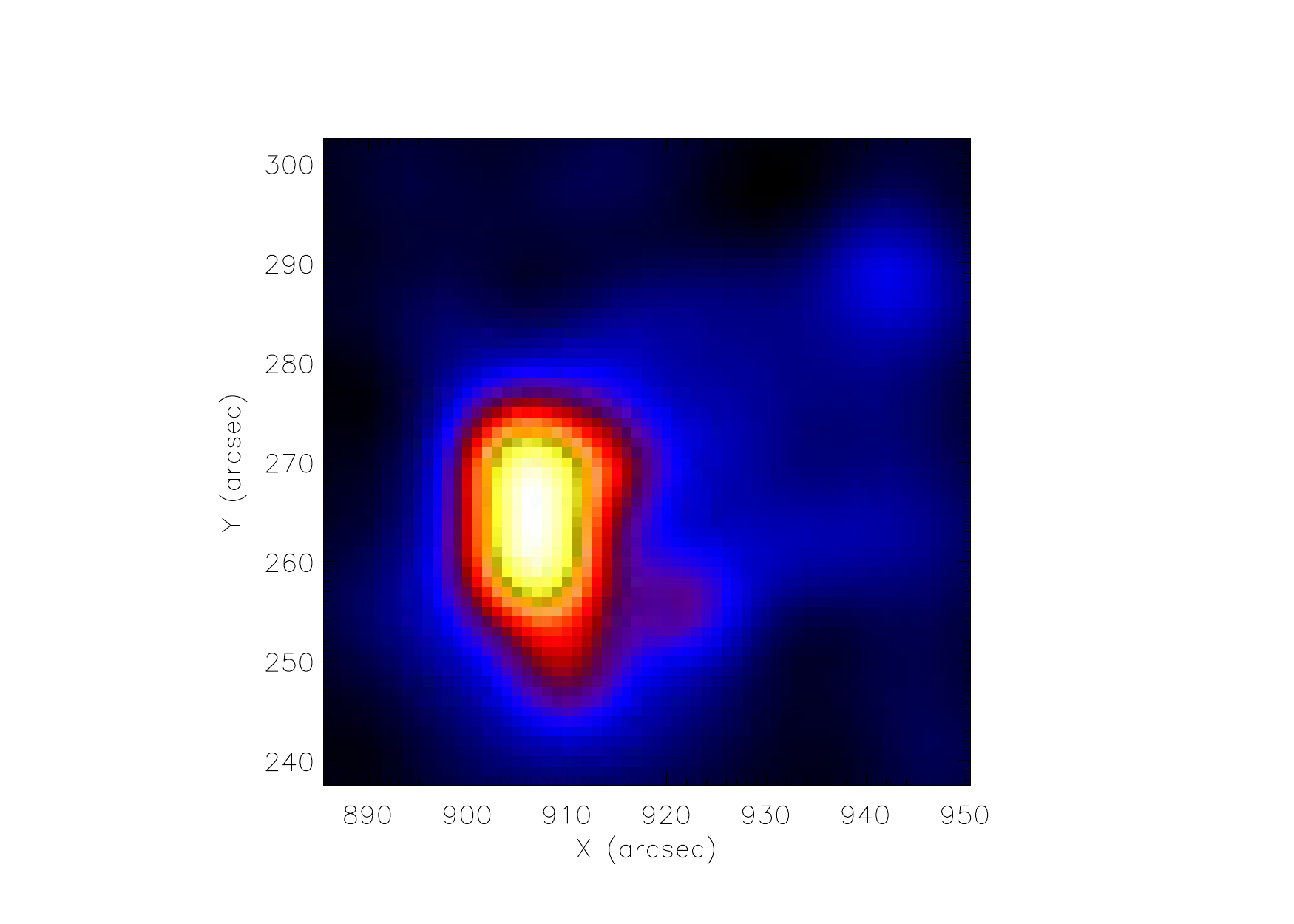}}
\subfloat[VFF]{\includegraphics[height=4.5cm]{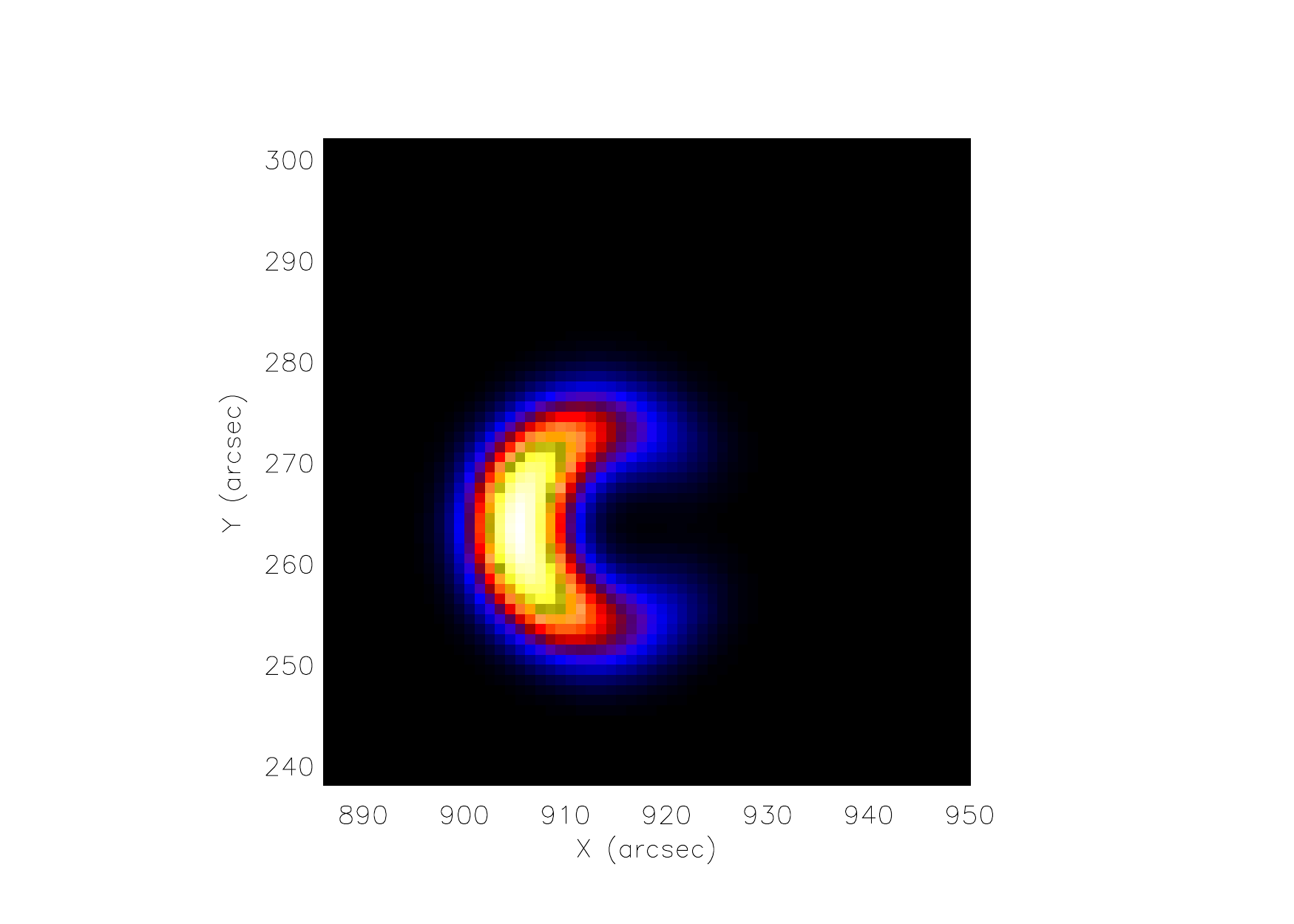}}
\subfloat[Bayes]{\includegraphics[height=4.5cm]{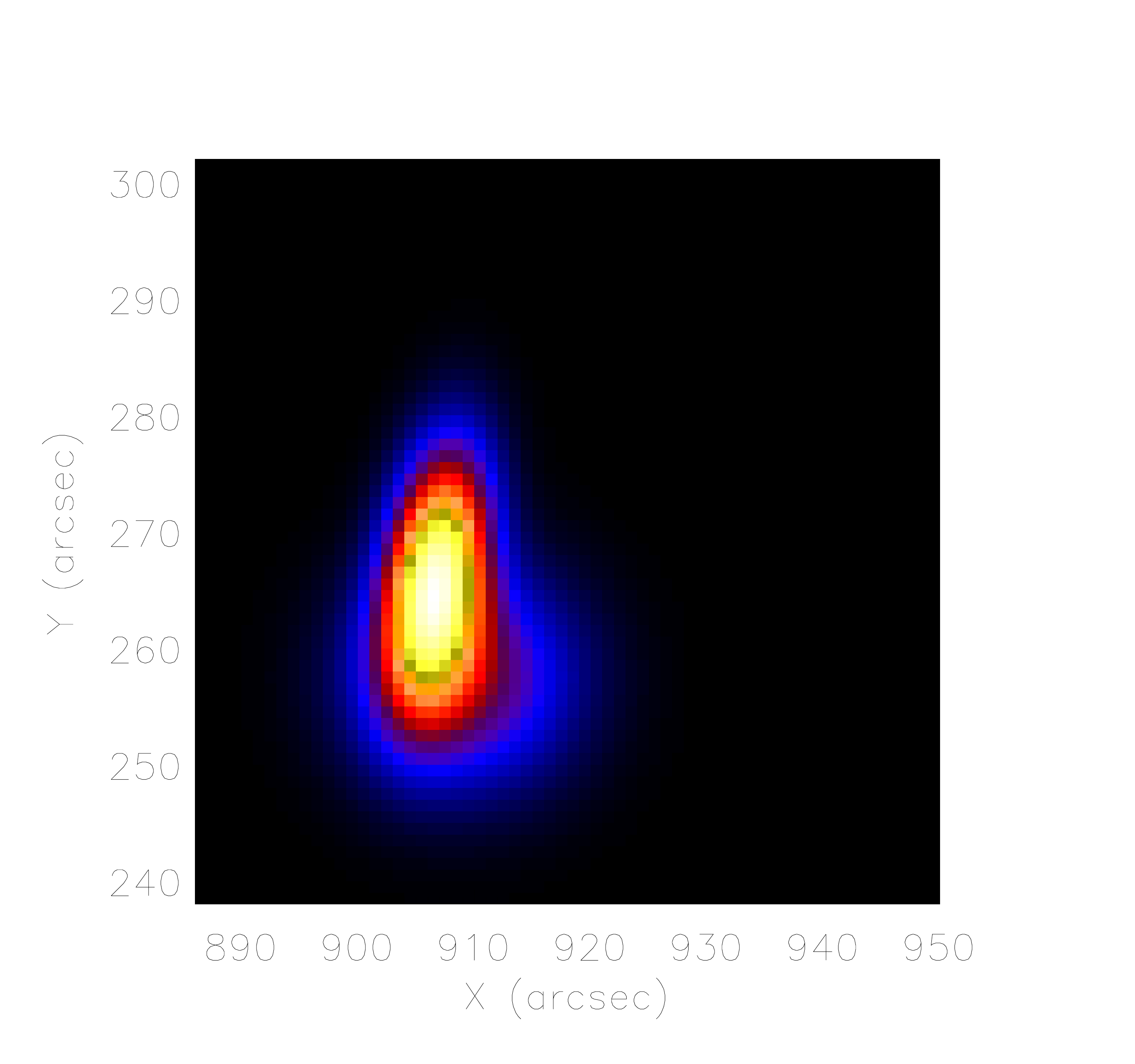}}
\caption{Images reconstructed in the 6-10 keV interval by Clean (left), forward fit (middle) and the Bayesian approach with $\lambda = 2$ and $p_C=1/2$, $p_E = p_L = 1/4$ (right).}
\label{fig:Bayes_image}
\end{center}
\end{figure*}

In Figure~\ref{fig:sui_num_sources} we show the posterior probability distribution of the number of sources in the six different energy bands. 
The distribution of the particles during the last iteration is shown in Figures~\ref{fig:sui_xy} and \ref{fig:sui_xy_mem}, together with the reconstructions obtained by Clean and MEM-NJIT, respectively; the values in the color bands represent the weights $\omega_i$ of each particle, with $\omega_i \in [0,1]$, with $i \in [1,N]$. Please notice that this plot of the Monte Carlo samples and weights only summarizes information on the posterior probability of the source locations, and not on the source size or strength; specifically, a source whose location is very certain will show up as a very small cloud of points, irrespective of the source size, while a source with an uncertain position will appear as a rather widespread cloud. Combining the information from Figures~\ref{fig:sui_num_sources} and~\ref{fig:sui_xy} we can draw some quantitative conclusions about the number and locations of sources in the map. For the lowest energy band (6-10~keV) the probability of having two sources is almost $90\%$ (Figure~\ref{fig:sui_num_sources}). At first glance, this result seems to 
run counter to the results of \cite{2002SoPh..210..245S}, where only a single source, which lies between the two footpoints, was described. However, the reconstruction provided by Clean shows a rather complex shape for that source, which cannot be well explained by a combination of simple geometric forms. The Bayesian approach addresses this problem by approximating the source with the combination of a circle and an ellipse characterized by very close (almost superimposed) locations (see Figure~\ref{fig:sui_xy}, top left panel and Figure~\ref{fig:Bayes_image}). As a confirmation of this, we note that if we increase the probability of having a loop instead of a circle or an ellipse in \eqref{eq:pr_sh}, the algorithm recovers a single loop-source with a significantly higher probability and with parameters and uncertainties close to the ones obtained by VFF (see Table \ref{tab:sui_6_10}). 
Thus, we can conclude that only a single source is found in the energy band 6-10 keV, which we interpret as a thermal coronal source situated at the apex of a magnetic structure connecting the footpoints.

\begin{table*}[!h]
\centering
\caption{Values of the source parameters provided by VFF and by the Bayesian method when the probability of having a loop source is increased to $1/2$ and conditioned to the case when a single loop is reconstructed. The $6-10$ keV channel is considered.} \label{tab:sui_6_10}
\begin{tabular}{c|cccccccc}
\hline
  $\,$ &  & $x$ & $y$ & $\phi$ & r  &  $\varepsilon$ & $\alpha$ & $\beta$\\
\hline\hline
 \multirow{2}{*}{Bayesian approach} &      Loop & 908.17 & 263.79  &  7784.65 & 14.05  &  0.74         &  182.0 & 106.7 \\
& Standard deviation & $0.44$ &   $    0.56$ & $ 367.62$ & $   0.88$  & $  0.04$ & $ 13.4$ & $   19.1$ \\
\hline
\multirow{2}{*}{Visibility forward fitting}& Loop & 908.78 &  263.82 & 7905.22 & 14.87  & 0.72 & -177.6  & 124.8  \\
& Standard deviation & 0.85 &  0.42  & 278.12 & 1.01   &   0.11   & 5.5  & 41.9 \\
\hline
\end{tabular}
\end{table*}

In the next energy band (10-14~keV), the map (Figure~\ref{fig:sui_xy}(b)) appears  to have possibly three chromospheric sources  (the one in the middle having a much smaller weight than the others), but by looking at the distribution on the number of sources in Figure~\ref{fig:sui_num_sources}, we see a higher probability of  three sources (the two footpoints and the loop-top source); the probability of having four sources present (red band in Figure~\ref{fig:sui_num_sources}) is around $3\%$. By comparing the weights of the particles in Figure~\ref{fig:sui_xy}, we see that most particles are at one of two footpoints, while only a few others constitute  the putative source in the middle of the two footpoints.

The map in the 14-20~keV energy band shows a predominance of three sources (Figure~\ref{fig:sui_num_sources}): Figure~\ref{fig:sui_xy}(c) shows these to be two footpoint sources plus the looptop source. The position of the southern footpoint is somewhat uncertain due to its relatively low intensity.

\begin{figure}[!h]
\begin{center}
\includegraphics[height=3cm]{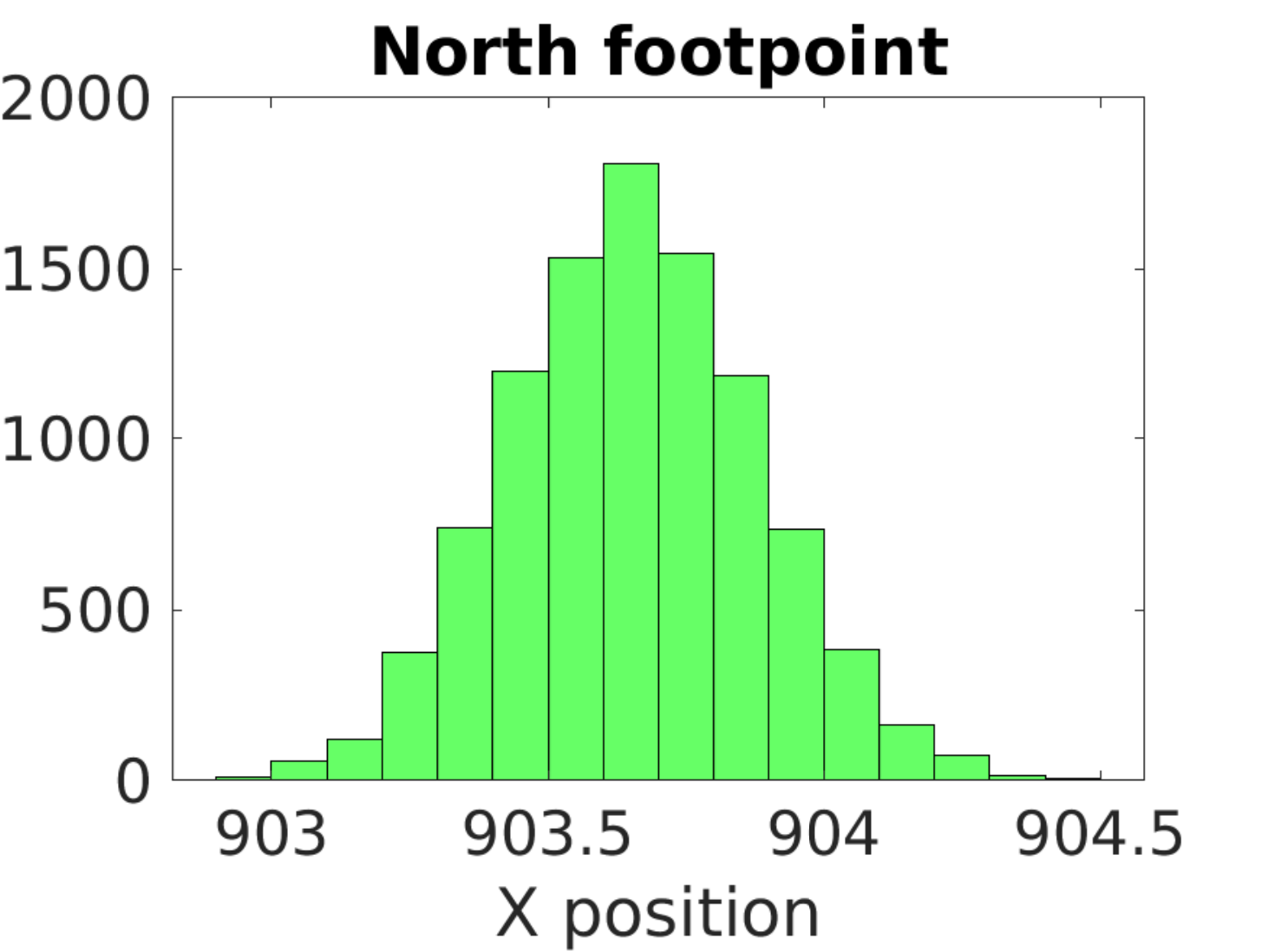}
\includegraphics[height=3cm]{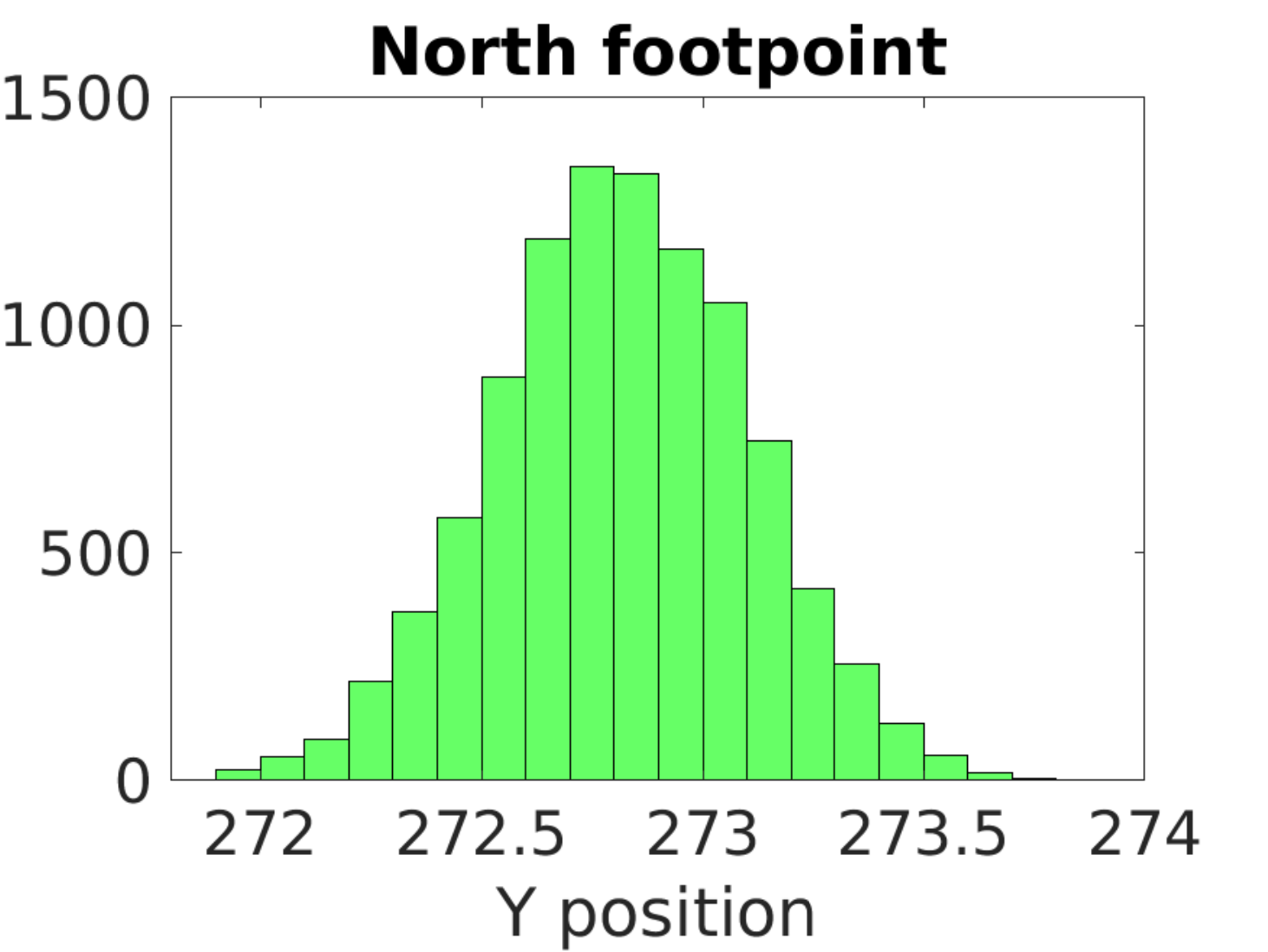}
\includegraphics[height=3cm]{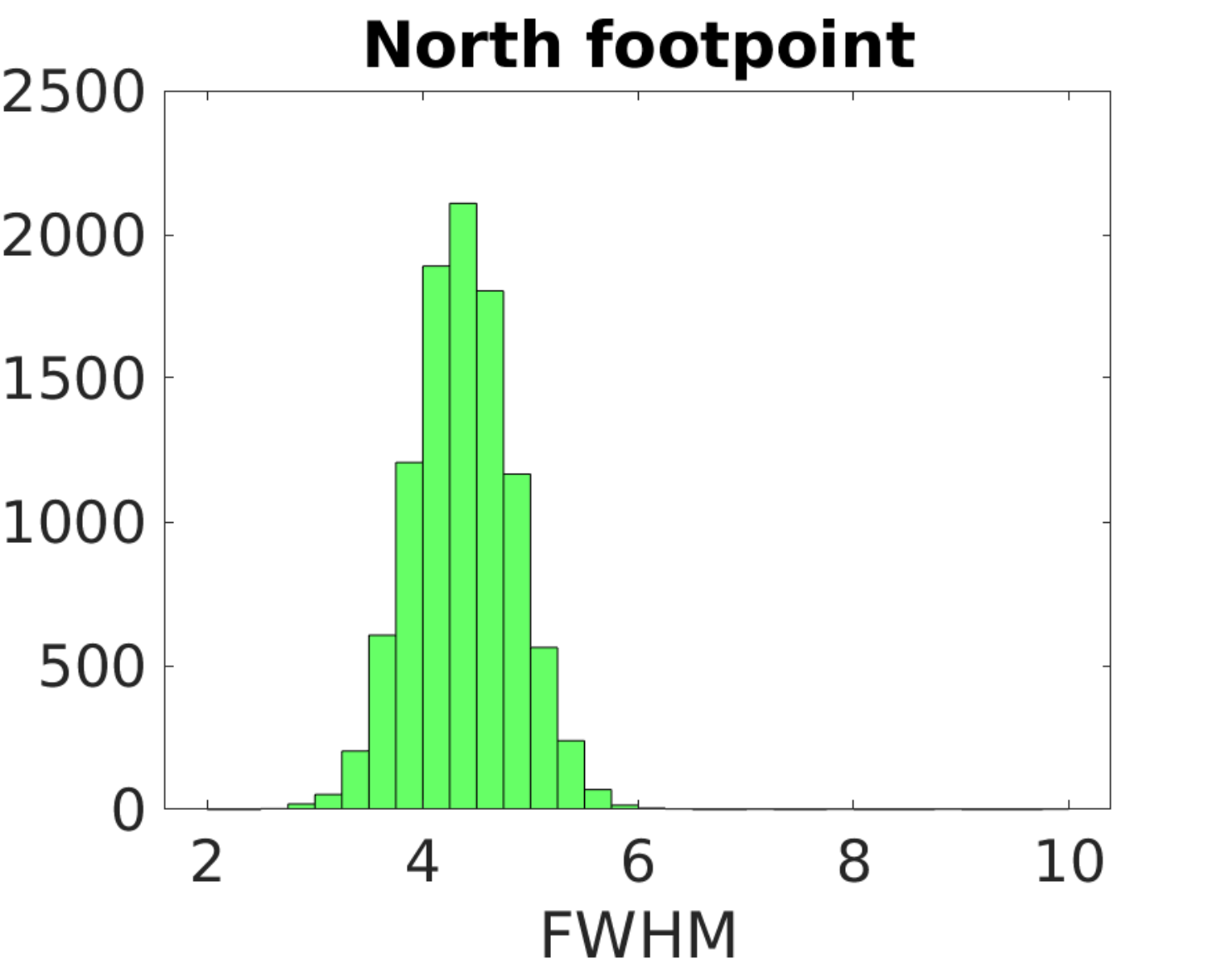}
\includegraphics[height=3cm]{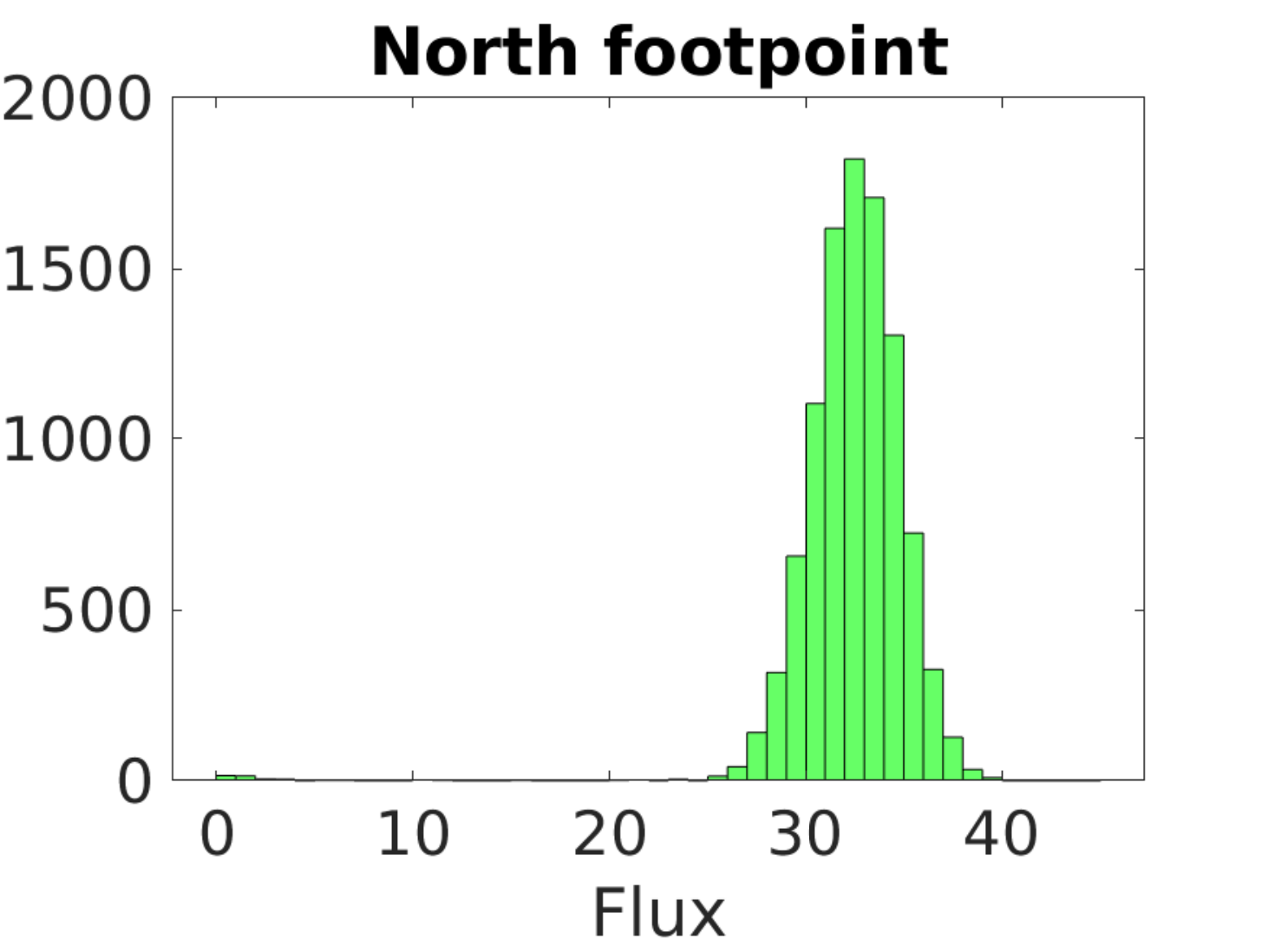}
\\
\includegraphics[height=3cm]{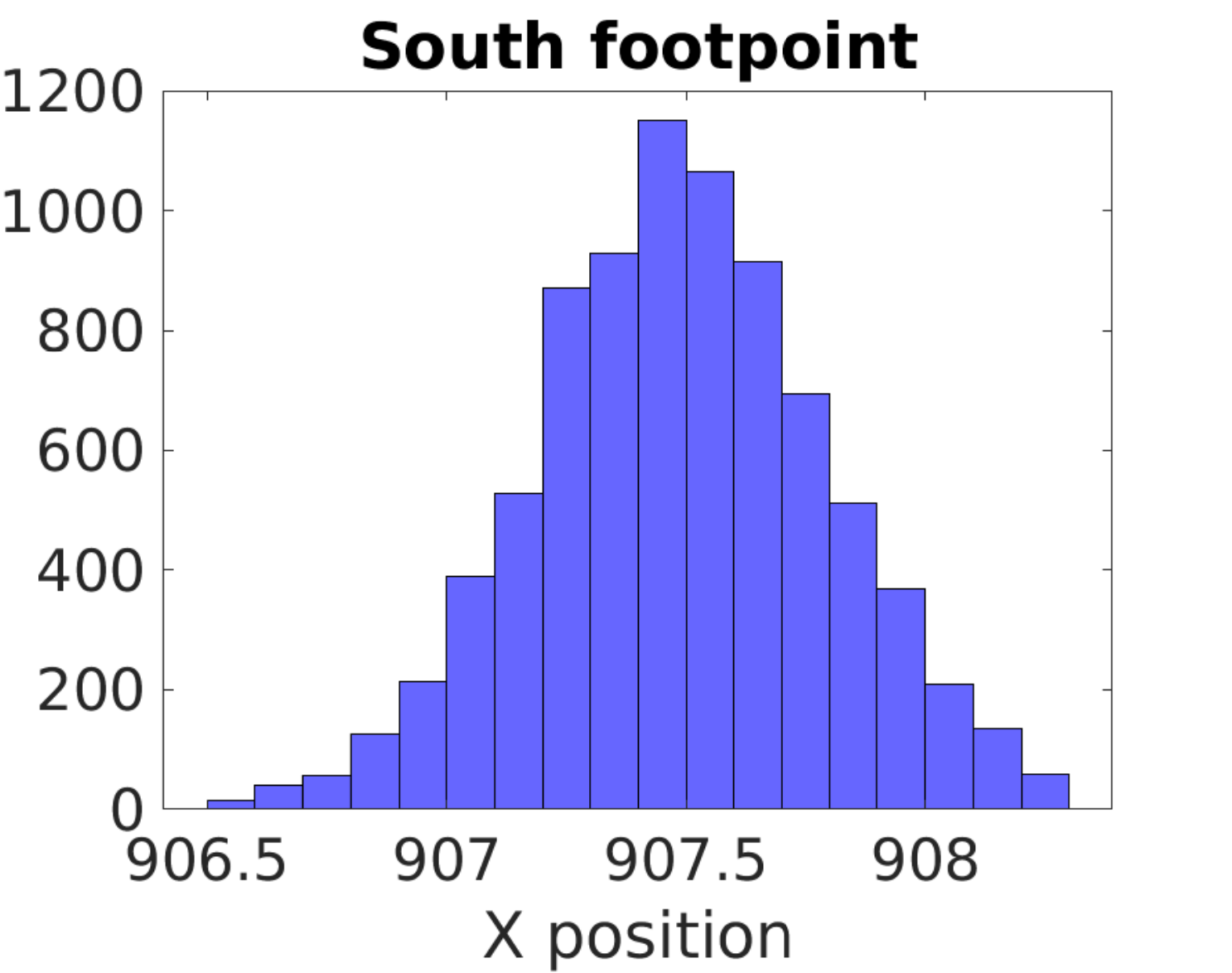}
\includegraphics[height=3cm]{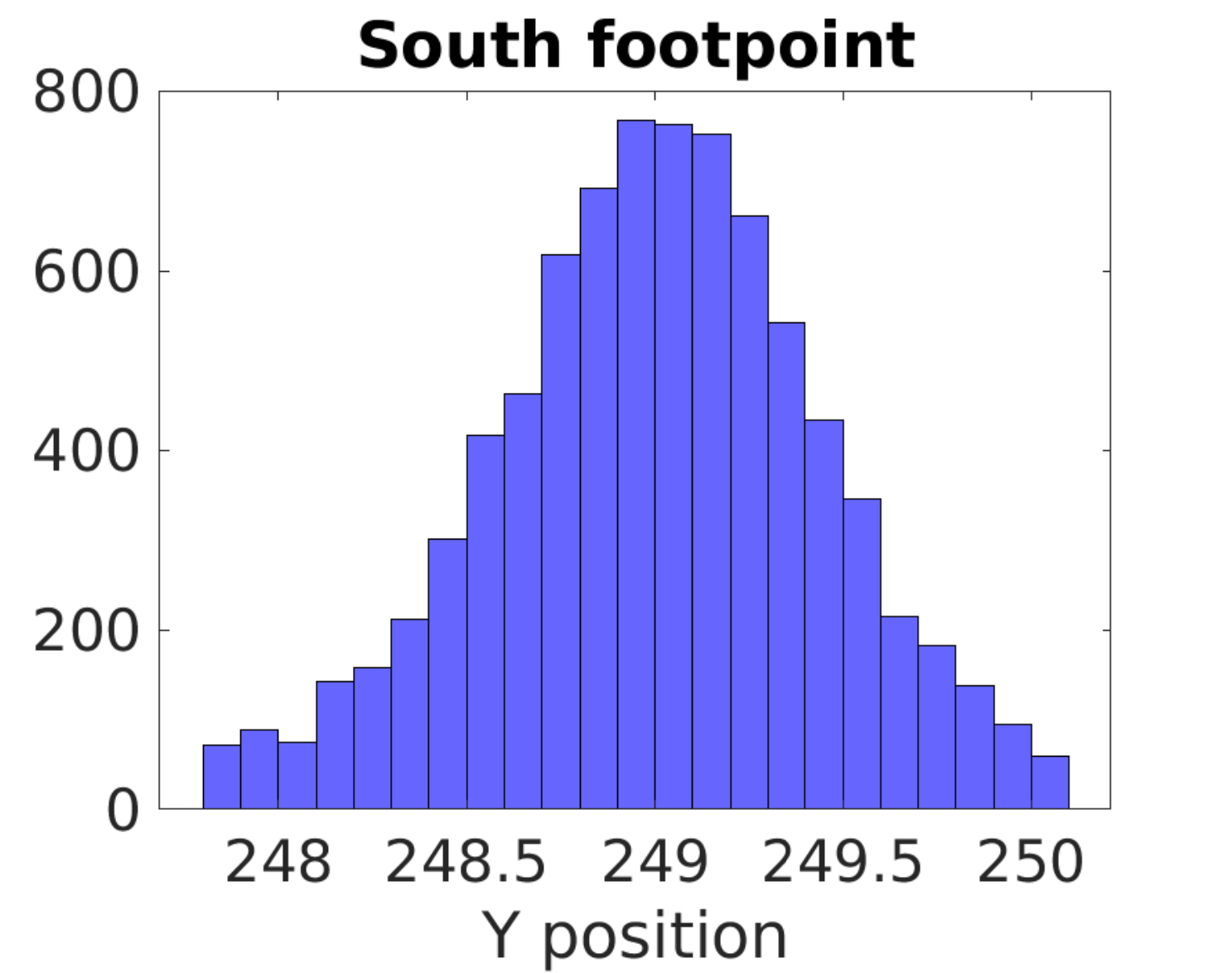}
\includegraphics[height=3cm]{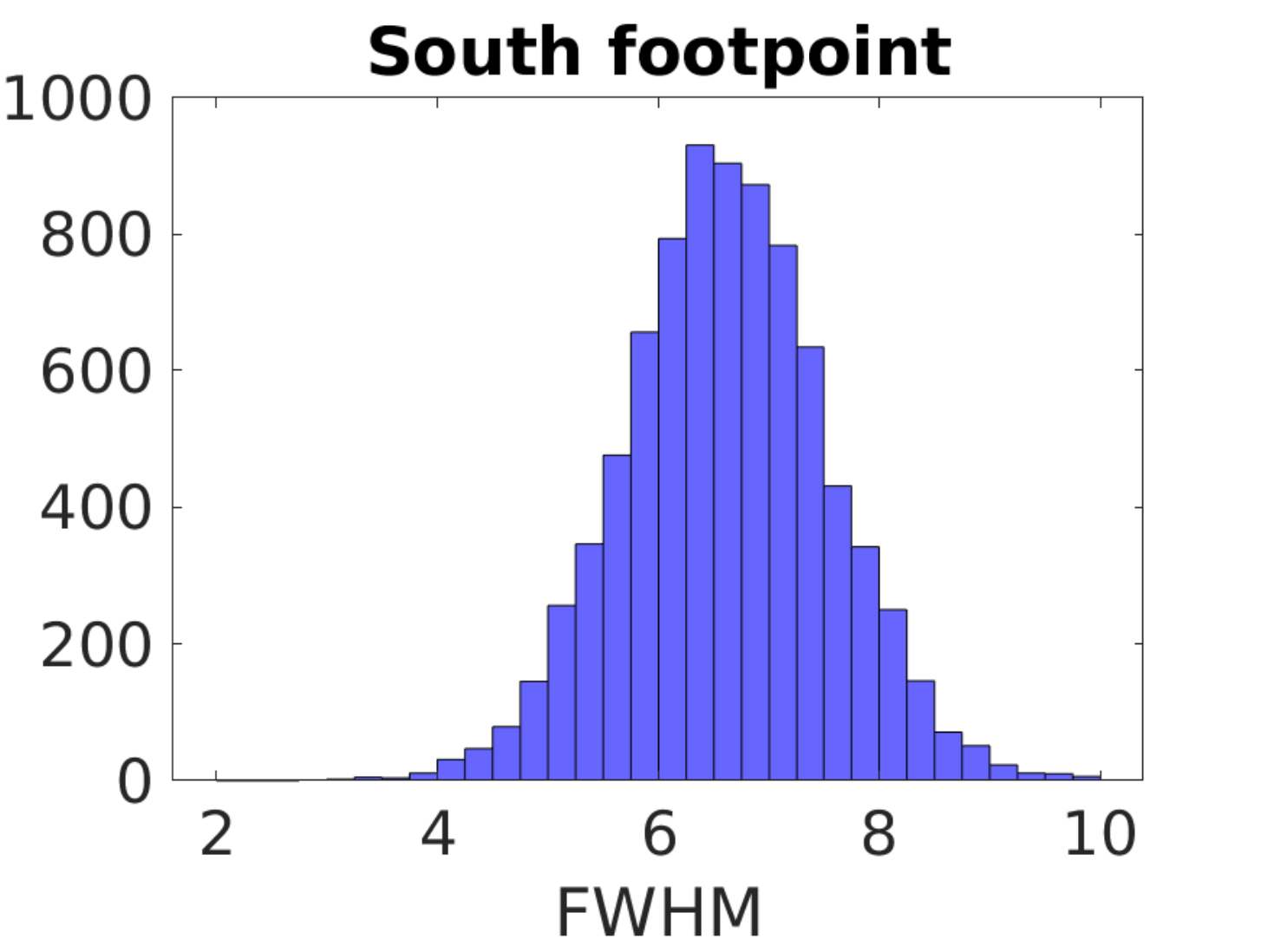}
\includegraphics[height=3cm]{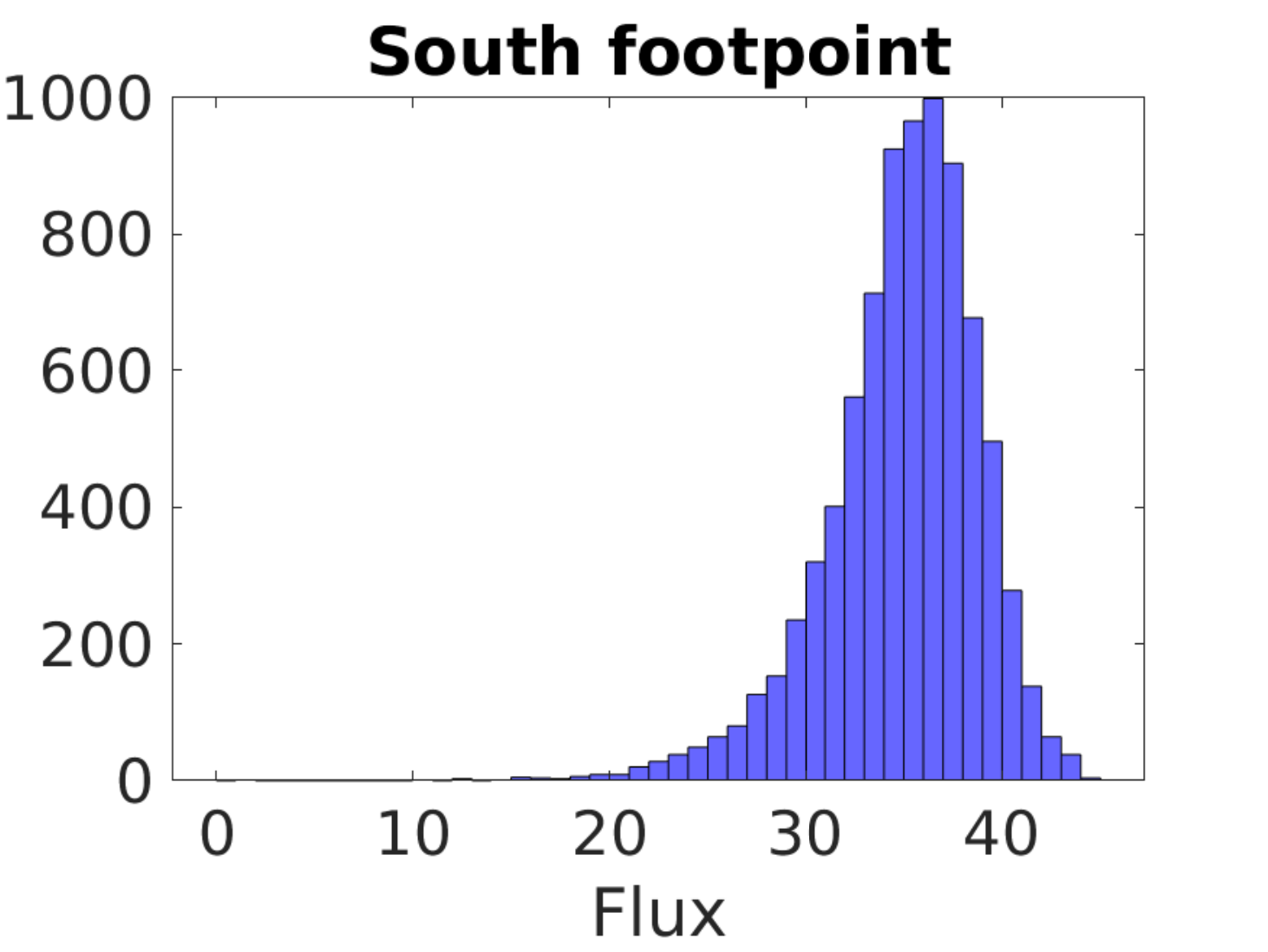}
\\
\includegraphics[height=3cm]{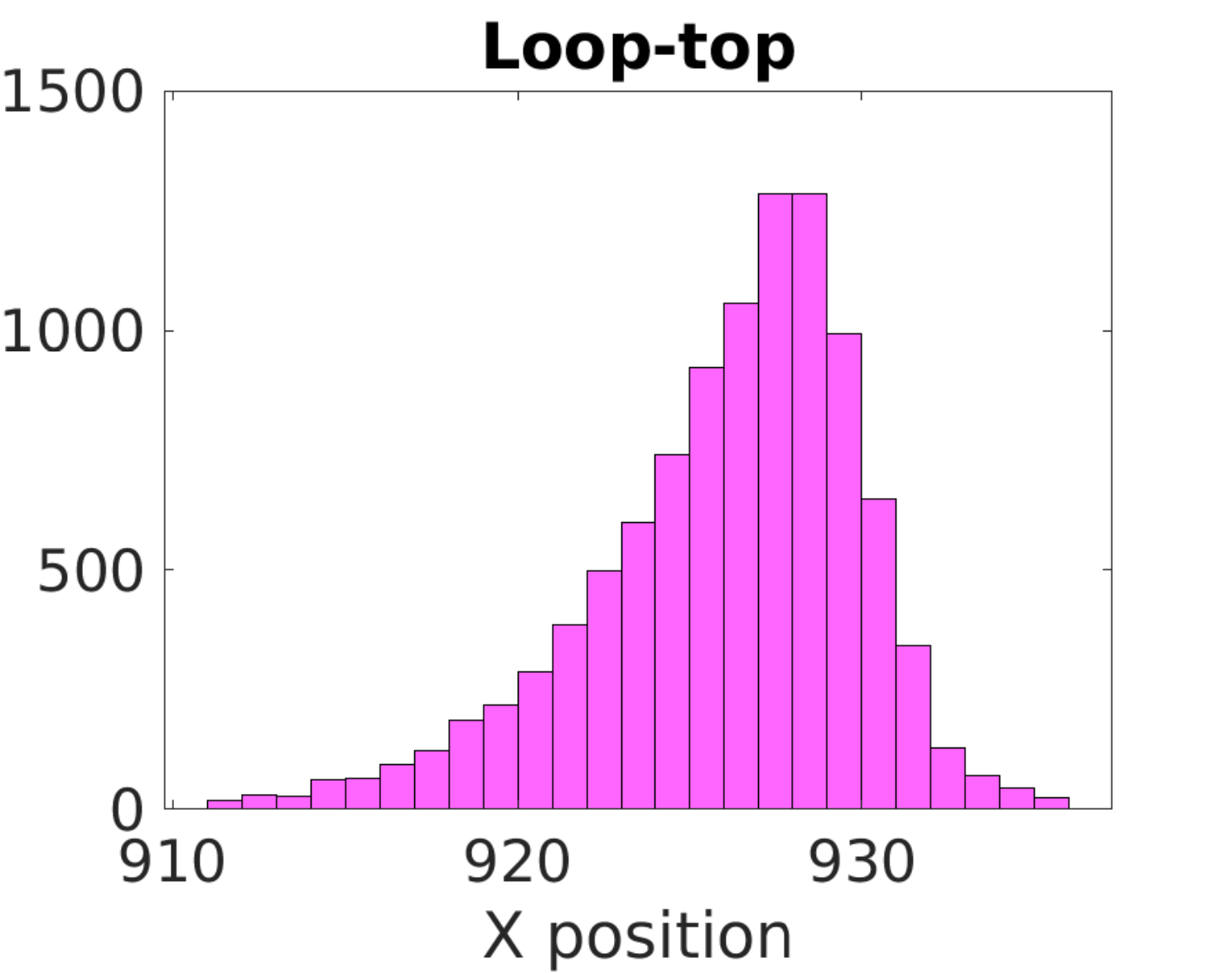}
\includegraphics[height=3cm]{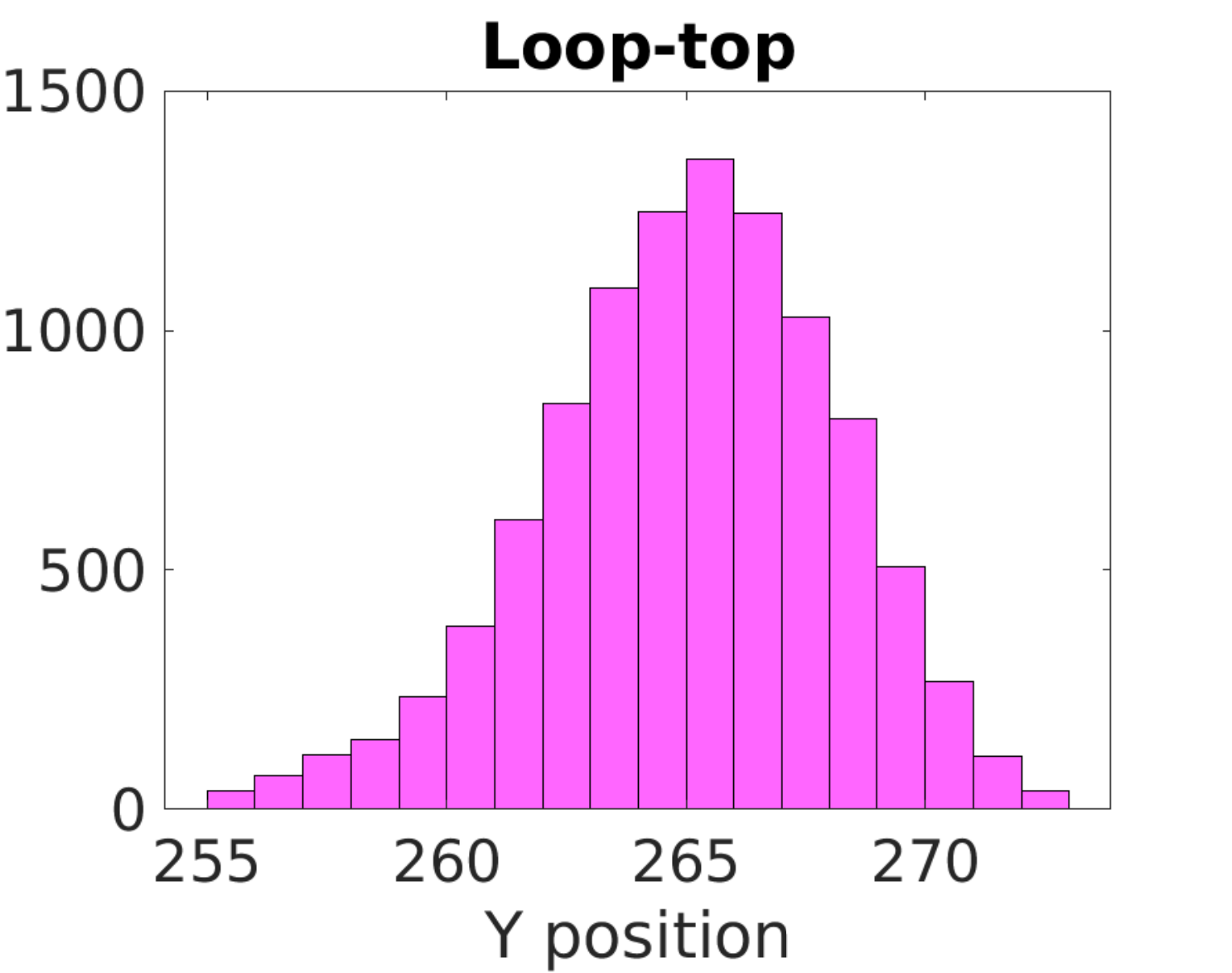}
\includegraphics[height=3cm]{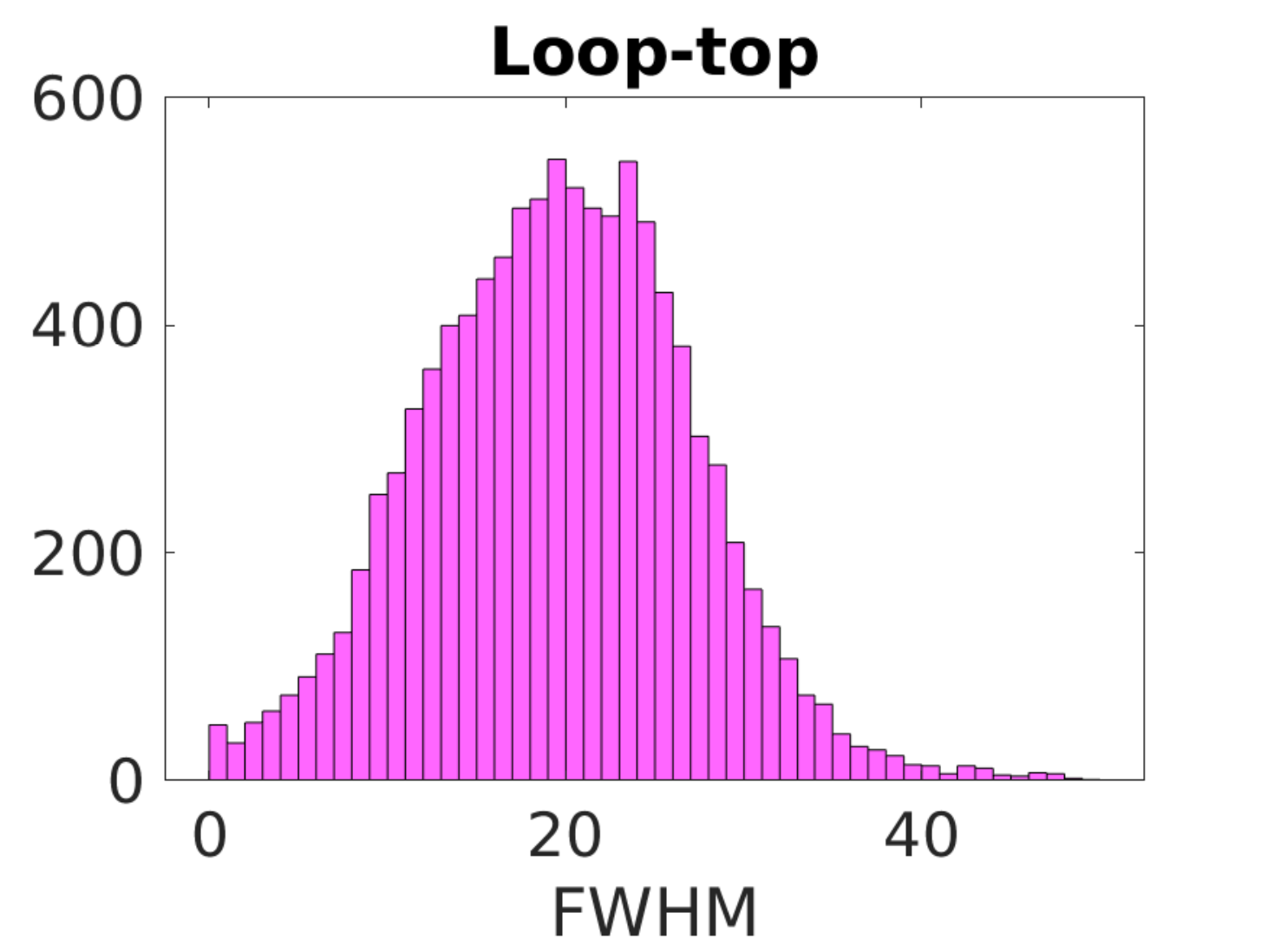}
\includegraphics[height=3cm]{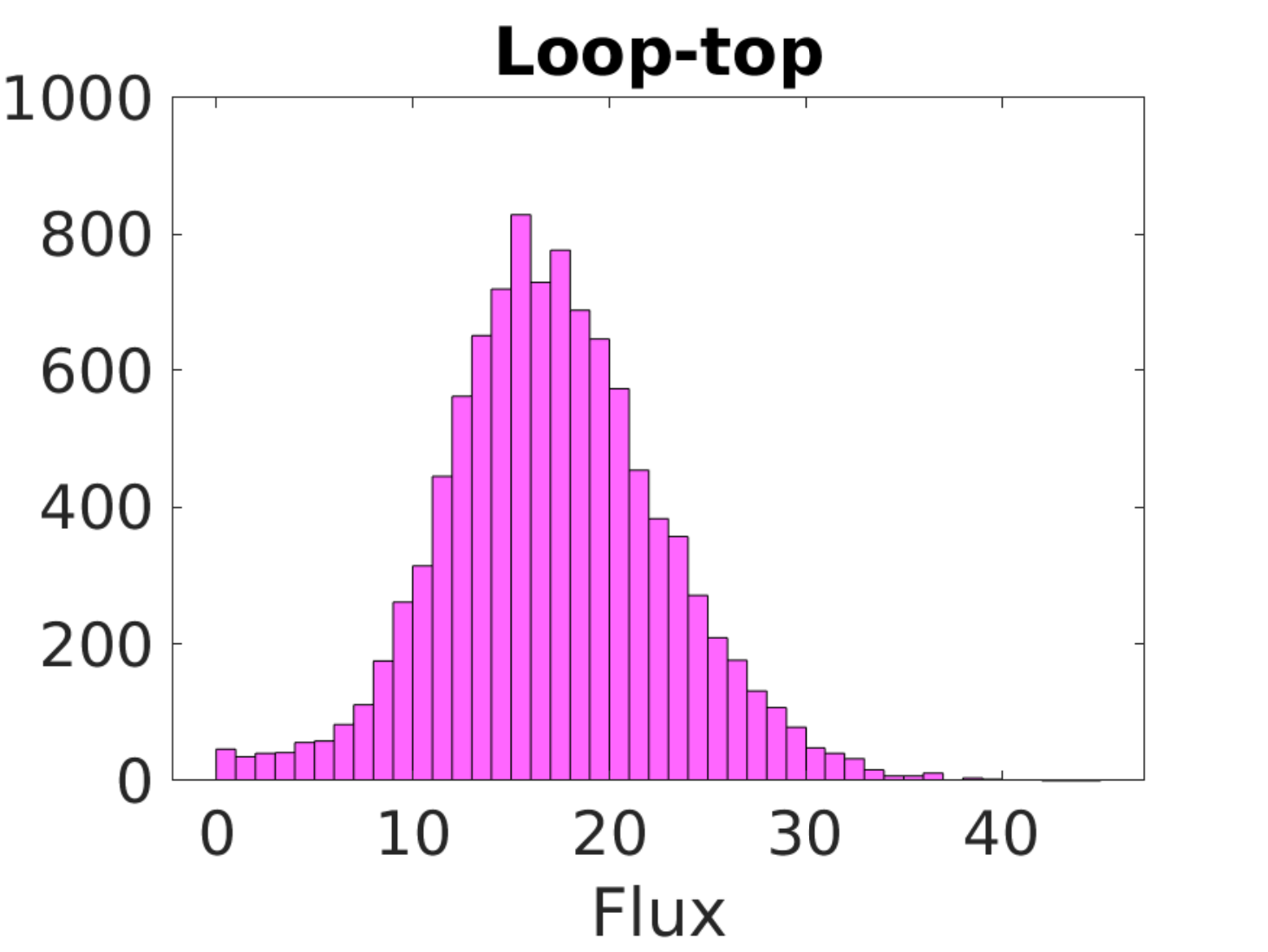}

\end{center}
\caption{Histograms of the Monte Carlo samples for the three sources in the energy band 20-30 keV; these histograms can be interpreted as (un-normalized) posterior distributions for the respective parameters.}
\label{fig:sui_param}
\end{figure}

The energy band from 20-30~keV is probably the one which describes best the flexibility of the Bayesian approach, with the highest probability of having three sources as found by \cite{2002SoPh..210..245S} and with a significant probability of four sources present, including a third footpoint, as found by \citet{2002SoPh..210..229K} (see next Section). The positions of the two main footpoints are completely identified, while the location of the loop-top source is a bit uncertain. In addition, our results suggest that the looptop source is relatively weak, which is coherent with the figures in \cite{2002SoPh..210..245S}, and relatively large. To show this, in Figure~\ref{fig:sui_param} we plot the histograms that represent the distribution of the positions, FWHM and flux for each source. In order to provide a good estimate of the FWHM of the smaller sources, we included in the analysis detectors 1 and 2, that take the spatial resolution to 2.26 arcsec.

For the next energy band (30-50~keV) the number of sources in the next case decreases, with a high probability, to two; only $24\%$ of the particles now pertain to the loop-top source.  The position of the (two) footpoints is clearly determined, while the third footpoint source is no longer clearly identified.  

Finally, for the last interval, 50-70~keV, the loop-top source has totally disappeared.  Only  less than $10\%$ of the particles constitute a third source, which is not even identified as the loop-top source but probably represents only noise.

\subsection{Comparison with the results of Krucker \& Lin}\label{sec:kru}

\begin{figure*}[!h]
\includegraphics[height=4.5cm]{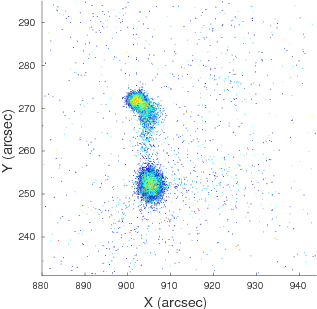}\hspace{0.5cm}
\includegraphics[height=4.5cm]{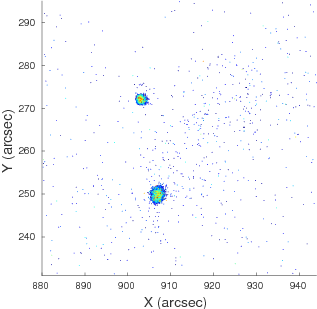}\hspace{0.5cm}
\includegraphics[height=4.5cm]{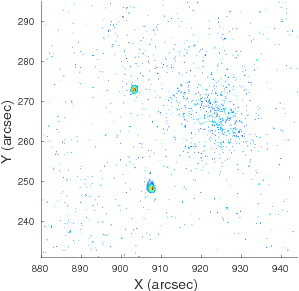}\\
\includegraphics[height=5cm]{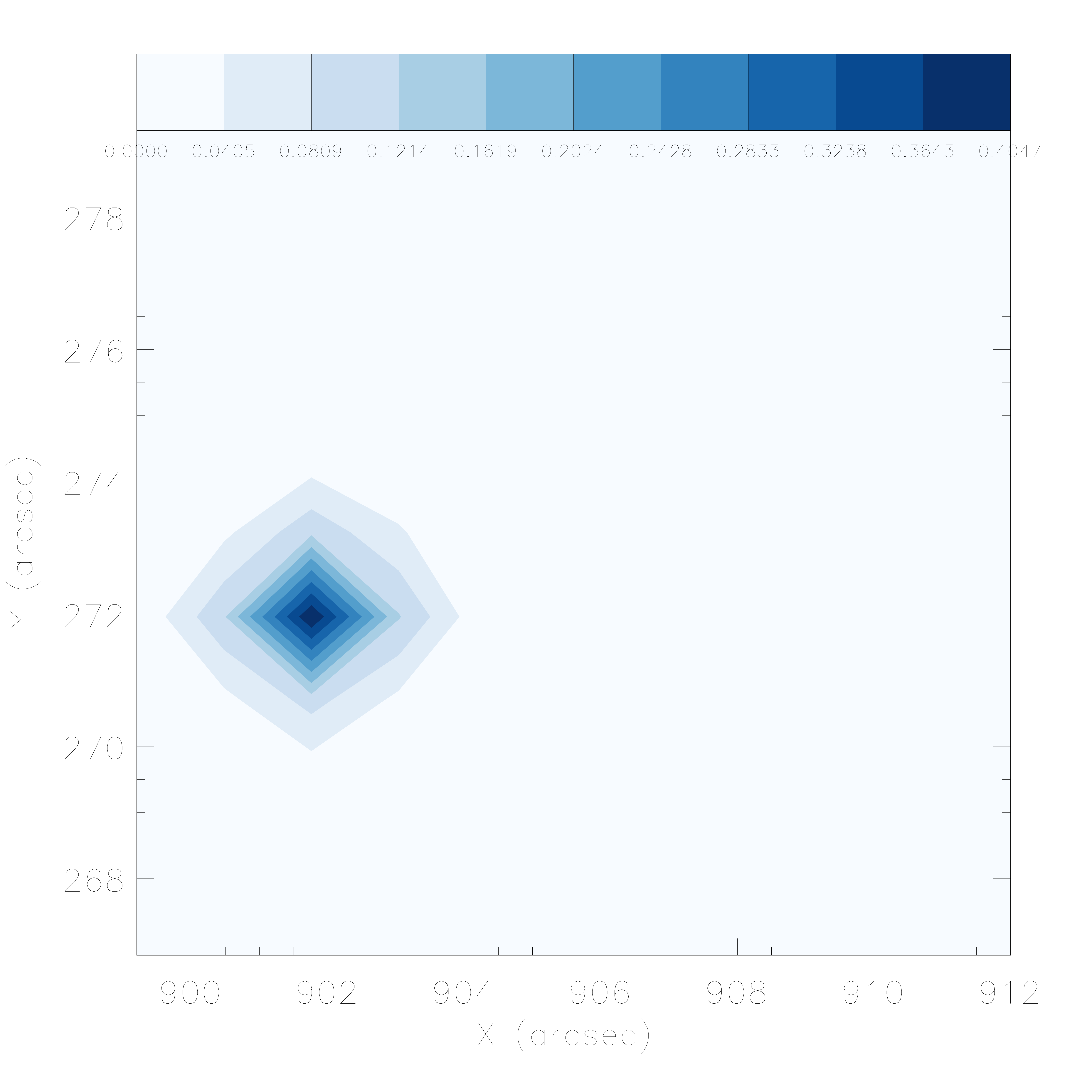}
\includegraphics[height=5cm]{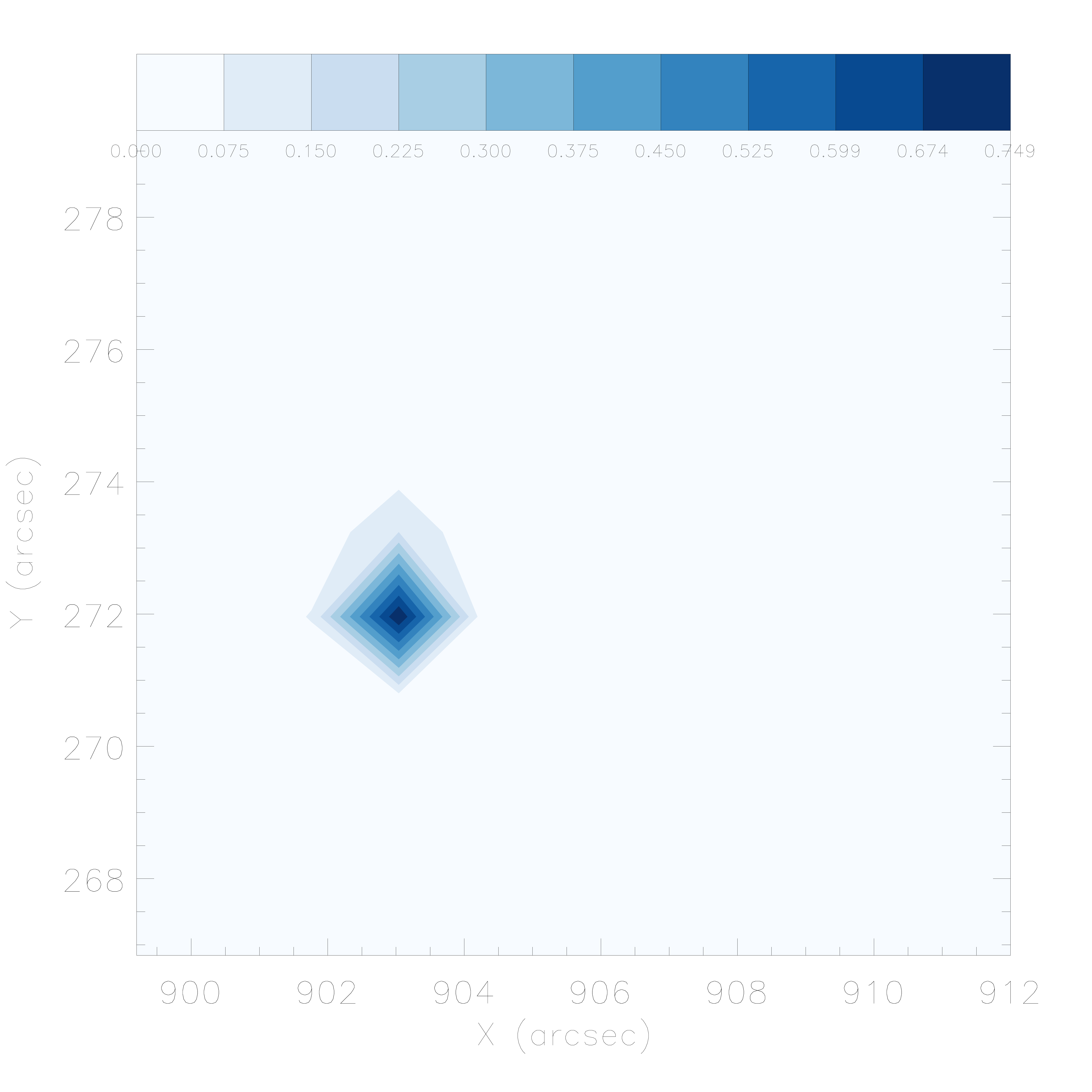}
\includegraphics[height=5cm]{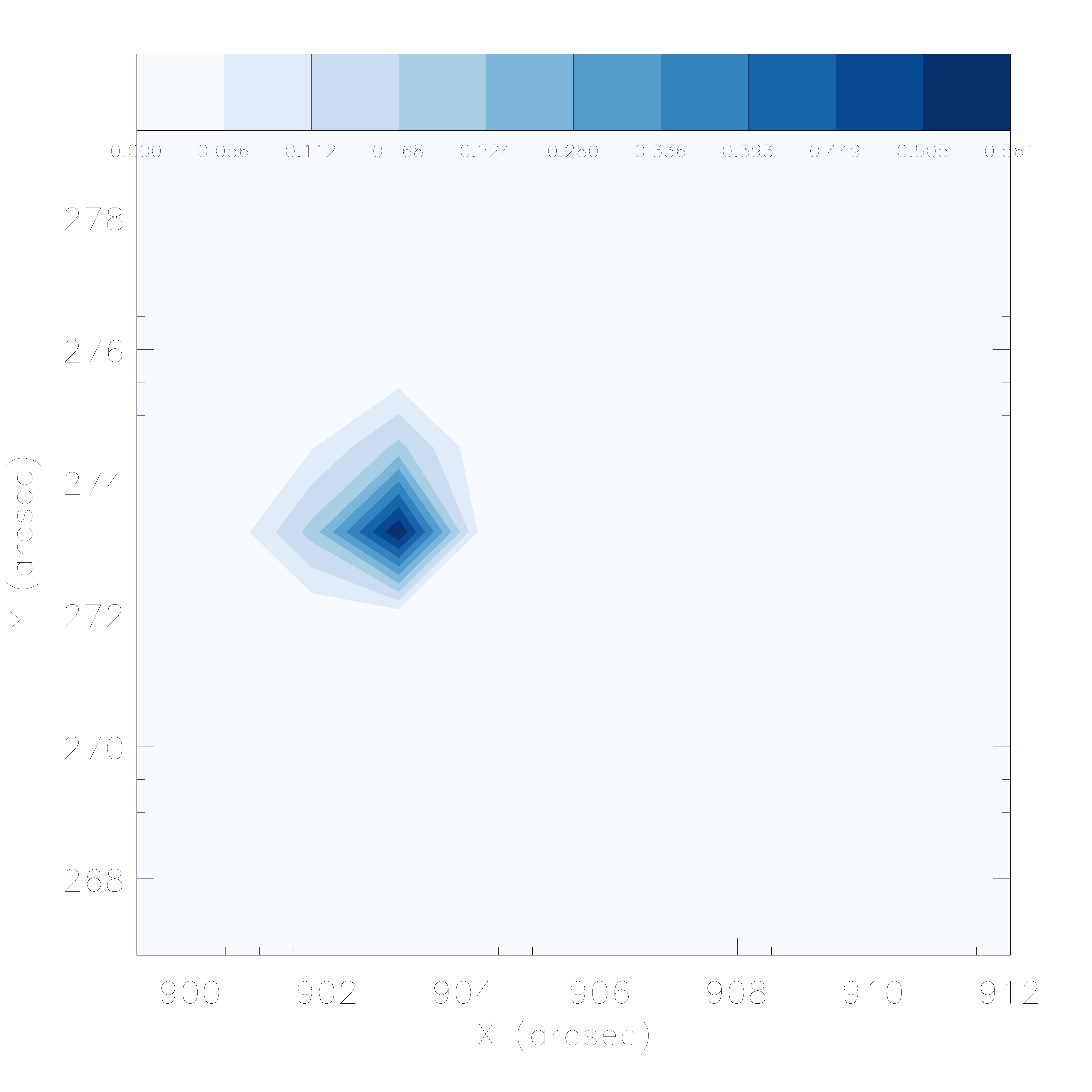}\\
\includegraphics[height=5cm]{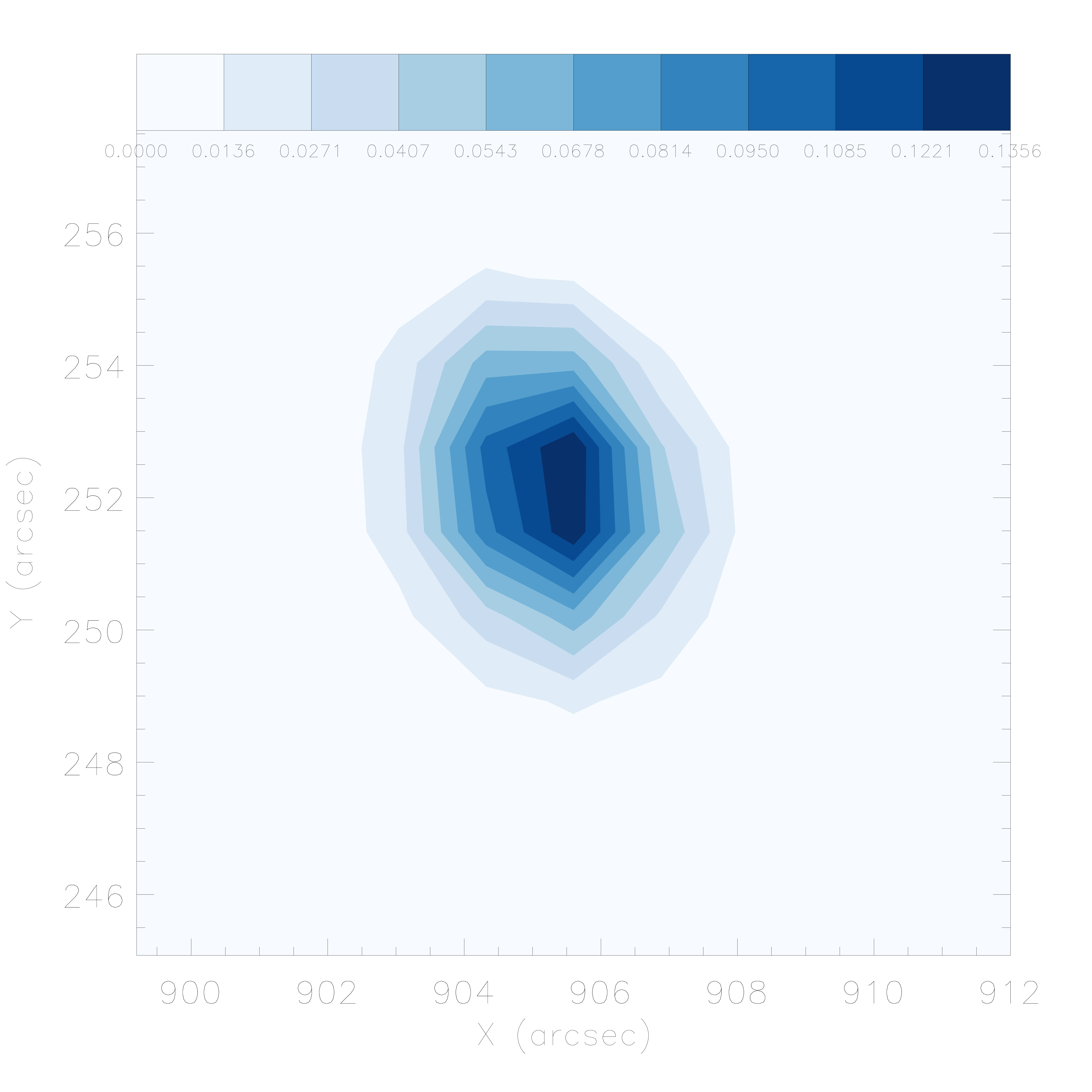}
\includegraphics[height=5cm]{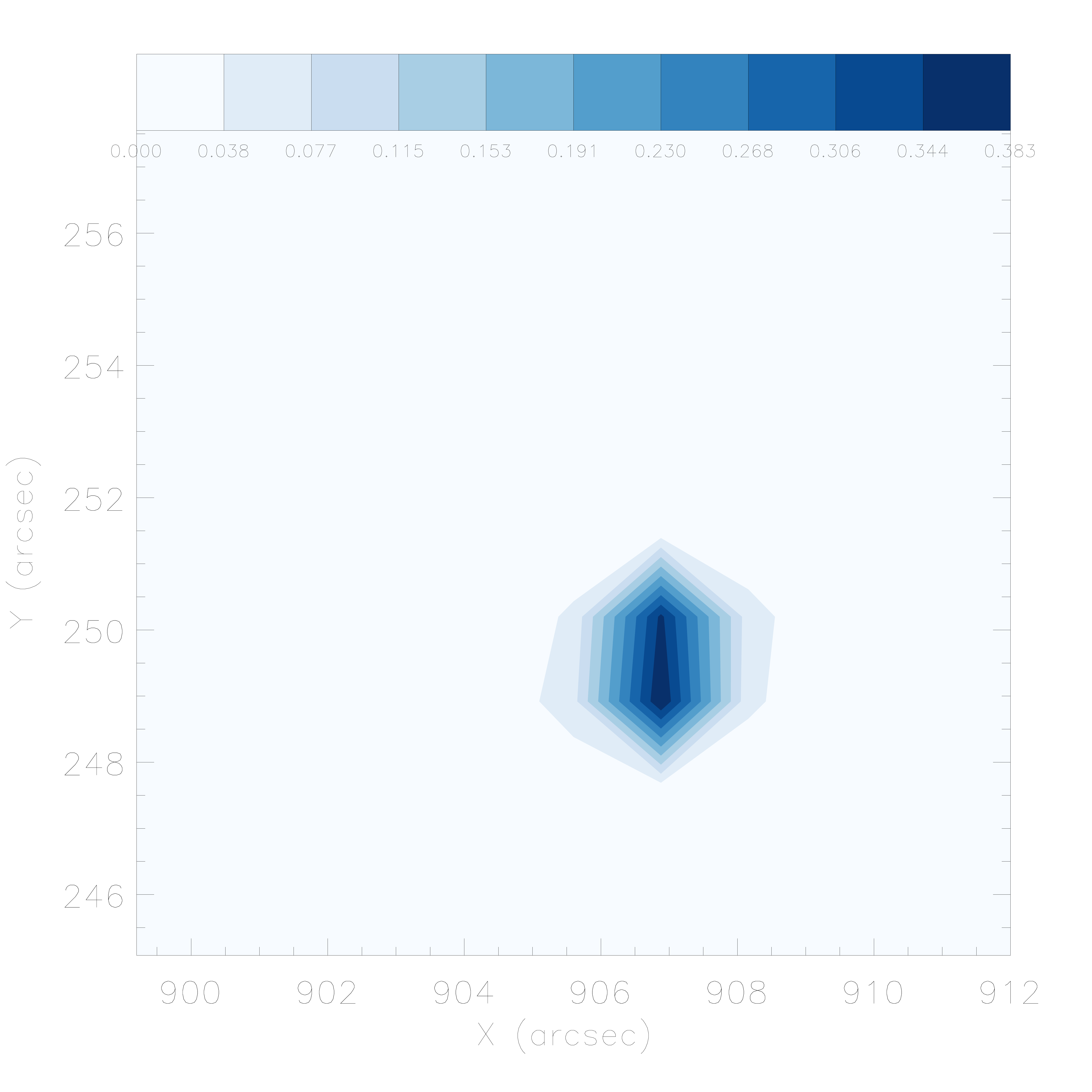}
\includegraphics[height=5cm]{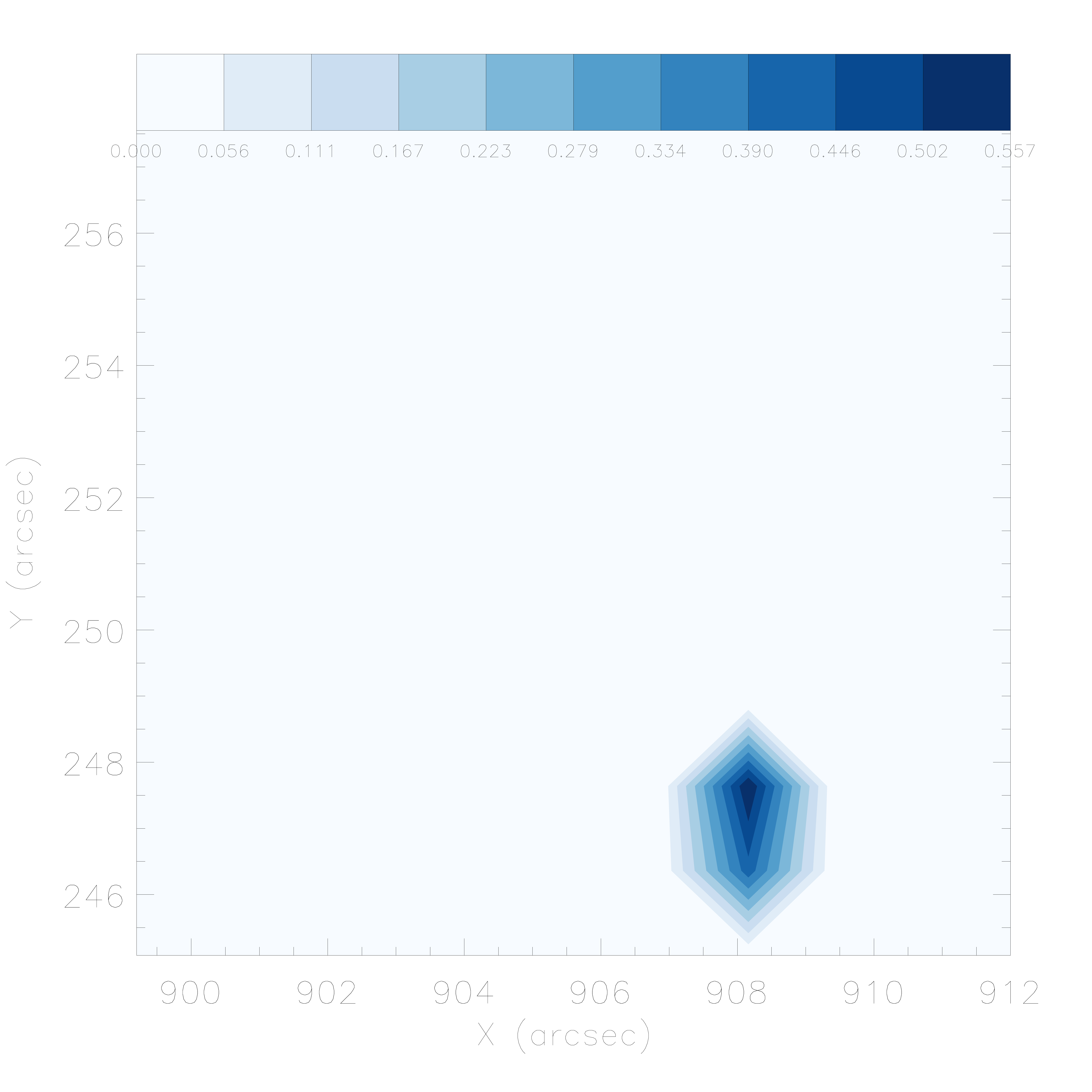}
\caption{Distribution of the particles (first row) for the 20 February flare at 11:06:08 UT, 11:06:15 UT, 11:06:24 UT (30-80~keV, 7'' resolution). The time intervals are $\sim 8$ s, $\sim 8$ s, and $\sim 16$ s, respectively. The histogram of the positions of northern and southern source  are shown in the second and third row, respectively. }
\label{fig:kru_new}
\end{figure*}

A similar analysis to the one in Section~\ref{sec:sui} was done for the results found in \cite{2002SoPh..210..229K}, which revealed a source structure with evolving complexity, including non-simultaneous brightening of different footpoints.  Figure~4 of \cite{2002SoPh..210..229K} shows that the Southernmost footpoint (``Source~2'') brightened some 8~s after the Northernmost one (``Source~1''); see also \cite{2002SoPh..210..261V}. Source~2 then slowly disappeared and a third source (``Source 3''), located more Southward than Source~2, reaches maximum brightness very shortly after the time of peak brightness of Source~2. To further explore these results, we first computed the source structure for the same three time intervals considered in Figure~4 of \citep{2002SoPh..210..229K}: 11:06:04-11:06:12, 11:06:11-11:06:19, and 11:06:16-11:06:32.  The first two intervals correspond to two full rotation periods of the \emph{RHESSI} spacecraft, while the third one corresponds to four rotations. In Figure~\ref{fig:kru_new}, the distributions of the particles at the last iteration and the histograms of the positions of the footpoints are shown. The regularization parameter has been set to $\lambda=3$ for all the three time intervals. Note that the particles in the first time interval (11:06:08 UT) are more spread than in the others two most probably due to the noise level in the data.  We can clearly see that the southern source has changed its coordinates, while the northern one is more stable. Table \ref{tab:kru_new} gives the values of the parameters  obtained by employing the proposed method and the VFF at different time intervals. It is worth noting that the publicly available version of the VFF algorithm works only with circular shapes in case of multiple sources. Therefore, with the VFF algorithm we cannot give an estimate of the eccentricity, rotation angle and loop angle. The parameter values are very similar for both recovering methods, the only difference regards the FWHM, due to the fact that the sources reconstructed by the VVF algorithm  are  circular. Clearly, we can conclude that for both methods the southern source is given by very close sources which appear at different locations and time intervals. Because these sources do not appear simultaneously, it seems likely to conclude that the southern source is a single footpoint moving towards south--southwest.

\begin{table*}[!h]
\centering
\caption{Values of the source parameters provided by VFF and by the Bayesian method for the three time intervals considered in Figure  \ref{fig:kru_new}  } \label{tab:kru_new}
\begin{tabular}{c|cccccccc|cccc}
\hline
\multirow{2}{*}{Time}  & \multicolumn{8}{c}{Bayesian method} & \multicolumn{4}{|c}{VFF} \\
&  $\,$ & $x$ & $y$ & $\phi$ & r  &  $\varepsilon$ & $\alpha$ & $\beta$ & $x$ & $y$ & $\phi$ & r\\
\hline\hline
\multirow{2}{*}{11:06:08 UT} &   Ellipse &  901.95  &  271.80  & 7.32   &   3.41  &    0.12    &   92.0    &     -0.4   &  902.58  &  271.30  & 11.82    &  5.53   \\
 &  Ellipse &  905.63  &  252.94  & 17.25   &   19.72  &    0.18    &  27.5    &     0.2   &  906.81  &  252.15   & 12.11    &   8.67  \\
\hline
\multirow{2}{*}{11:06:15 UT} & Circle & 903.00 &   271.96   & 24.20   &   4.03    &   0.10   &     22.8   & 0.1  & 903.19 &   271.99   & 25.13   &   3.99      \\
&Ellipse & 906.99 &   249.38   & 25.18  &   8.39    &   0.17  &    40.6  & -0.2       & 907.14 &   249.53  & 24.94  &  7.80 \\
\hline
\multirow{2}{*}{11:06:24 UT} & Ellipse & 902.95  &  273.22   & 14.39   &   4.68   &     0.19     & 46.3 &  0.2  &  902.89  &  273.16    & 14.96  &   4.67    \\
& Ellipse & 908.15 &   247.60 &24.27 & 6.20  & 0.30 & 66.2 & -0.0  &  908.04  &   247.60 & 24.96 & 6.18 \\
\hline
\end{tabular}
\end{table*}

\begin{figure}[!h]
\begin{center}
\subfloat[11:06:05]{\includegraphics[height=3.6cm]{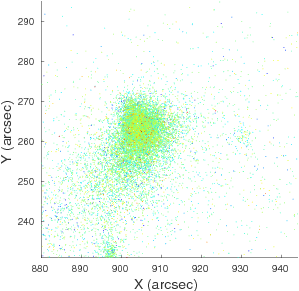}}
\subfloat[11:06:07]{\includegraphics[height=3.6cm]{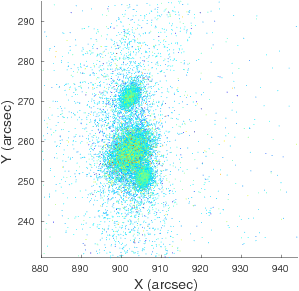}}
\subfloat[11:06:09]{\includegraphics[height=3.6cm]{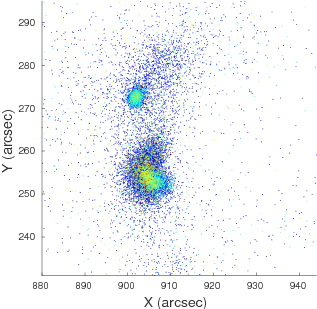}}
\subfloat[11:06:11]{\includegraphics[height=3.6cm]{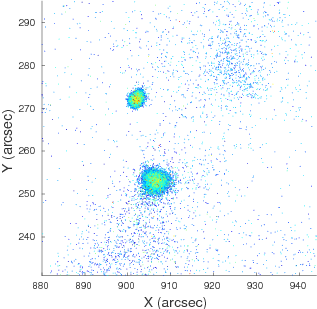}}
\subfloat[11:06:13]{\includegraphics[height=3.6cm]{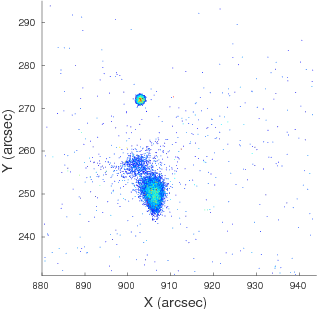}}\\
\subfloat[11:06:15]{\includegraphics[height=3.6cm]{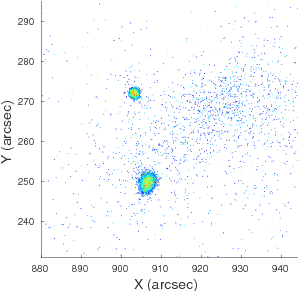}}
\subfloat[11:06:17]{\includegraphics[height=3.6cm]{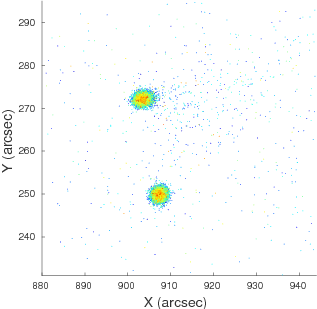}}
\subfloat[11:06:19]{\includegraphics[height=3.6cm]{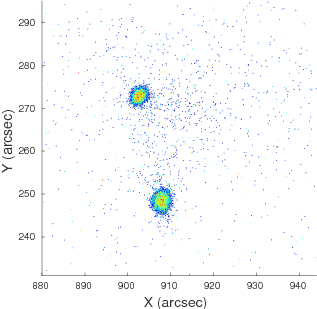}}
\subfloat[11:06:21]{\includegraphics[height=3.6cm]{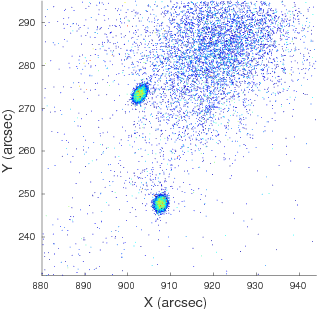}}
\subfloat[11:06:23]{\includegraphics[height=3.6cm]{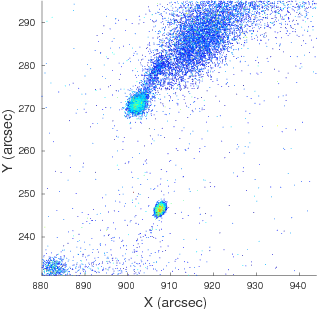}}\\
\subfloat[11:06:25]{\includegraphics[height=3.6cm]{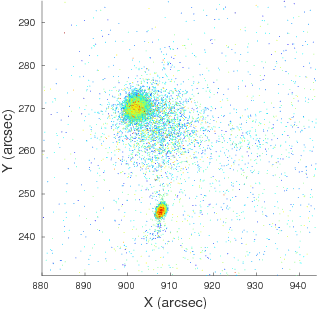}}
\subfloat[11:06:27]{\includegraphics[height=3.6cm]{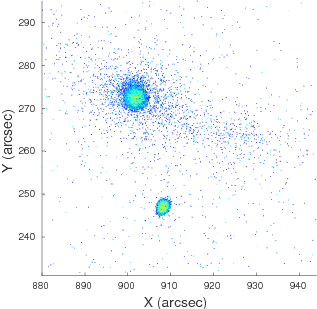}}
\subfloat[11:06:29]{\includegraphics[height=3.6cm]{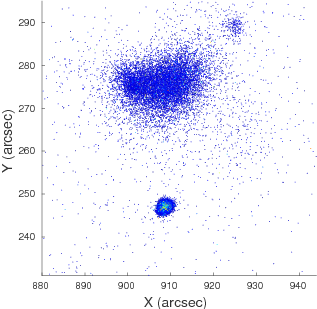}}
\subfloat[11:06:31]{\includegraphics[height=3.6cm]{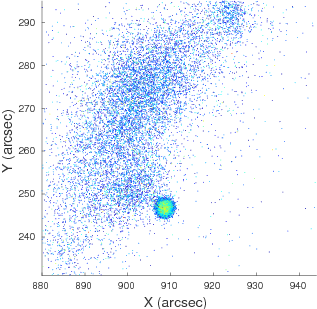}}
\subfloat[11:06:33]{\includegraphics[height=3.6cm]{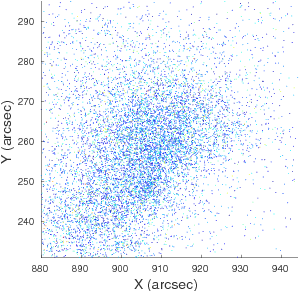}}
\end{center}
\caption{Distribution of the particles obtained by applying the Bayesian approach to visibilities taken every 2 s, with ∼ 4 s integration time. Energy interval: 30-80 keV, 7'' resolution. All times in UT. The parameter $\lambda$ has been set to 2 for all the time intervals. }
\label{fig:dynamic}
\end{figure}

\begin{figure}[!h]
\begin{center}
\includegraphics[height=4.1cm]{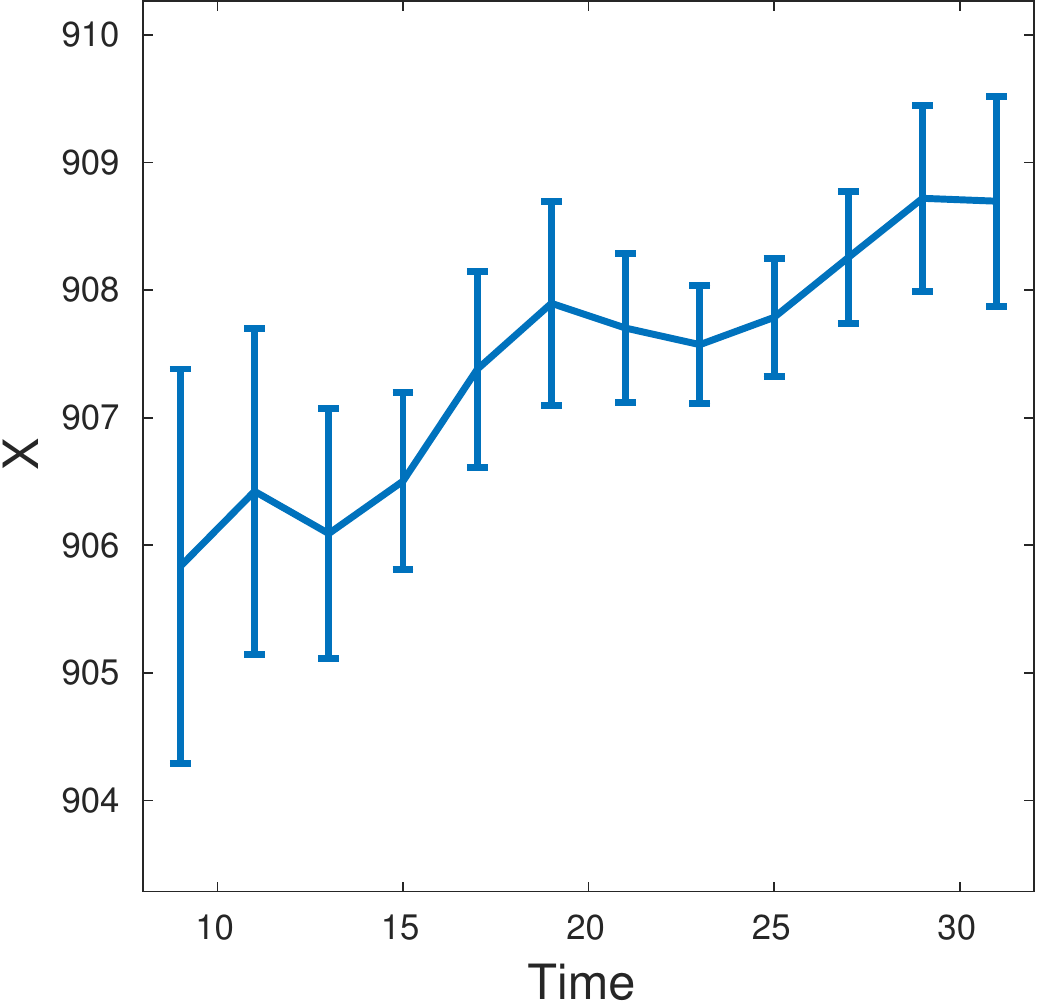}\hspace{0.2cm}
\includegraphics[height=4.1cm]{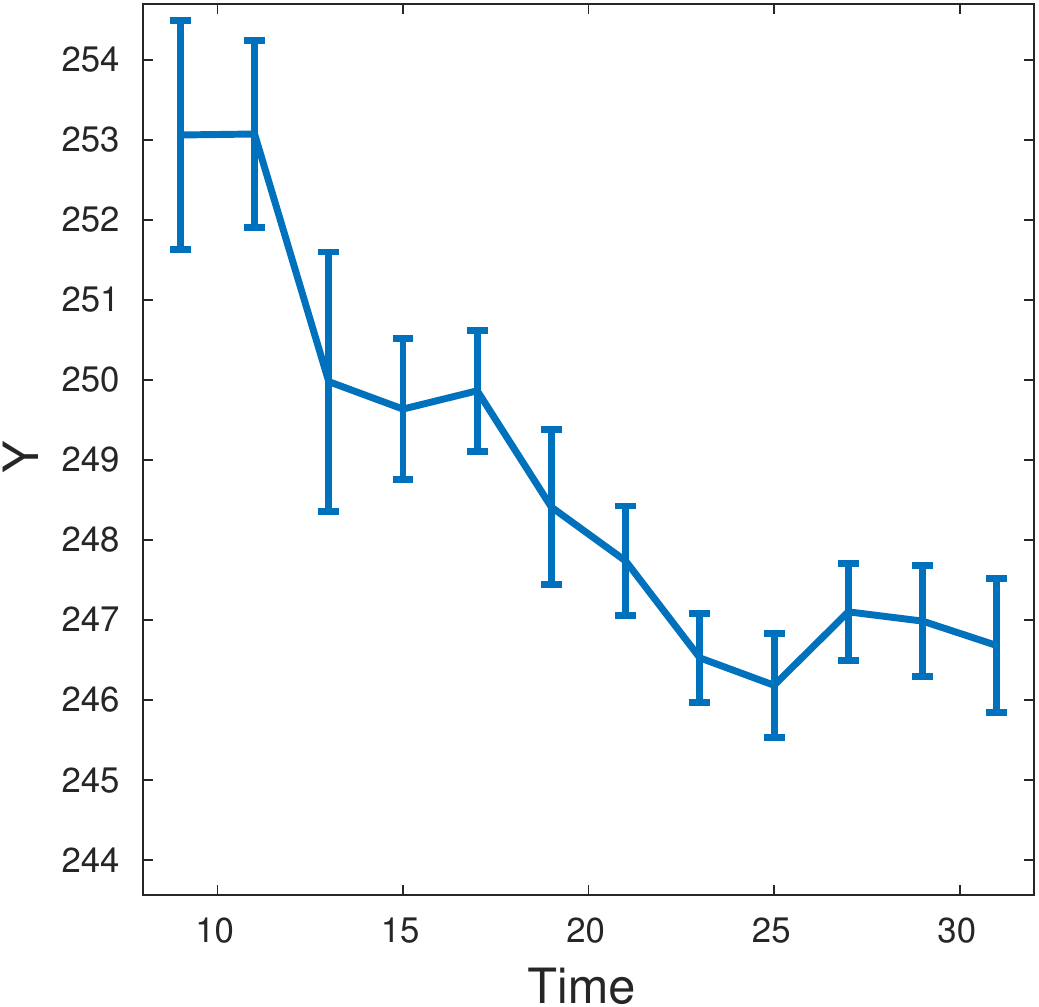}
\hspace{0.2cm}
\includegraphics[height=4.1cm]{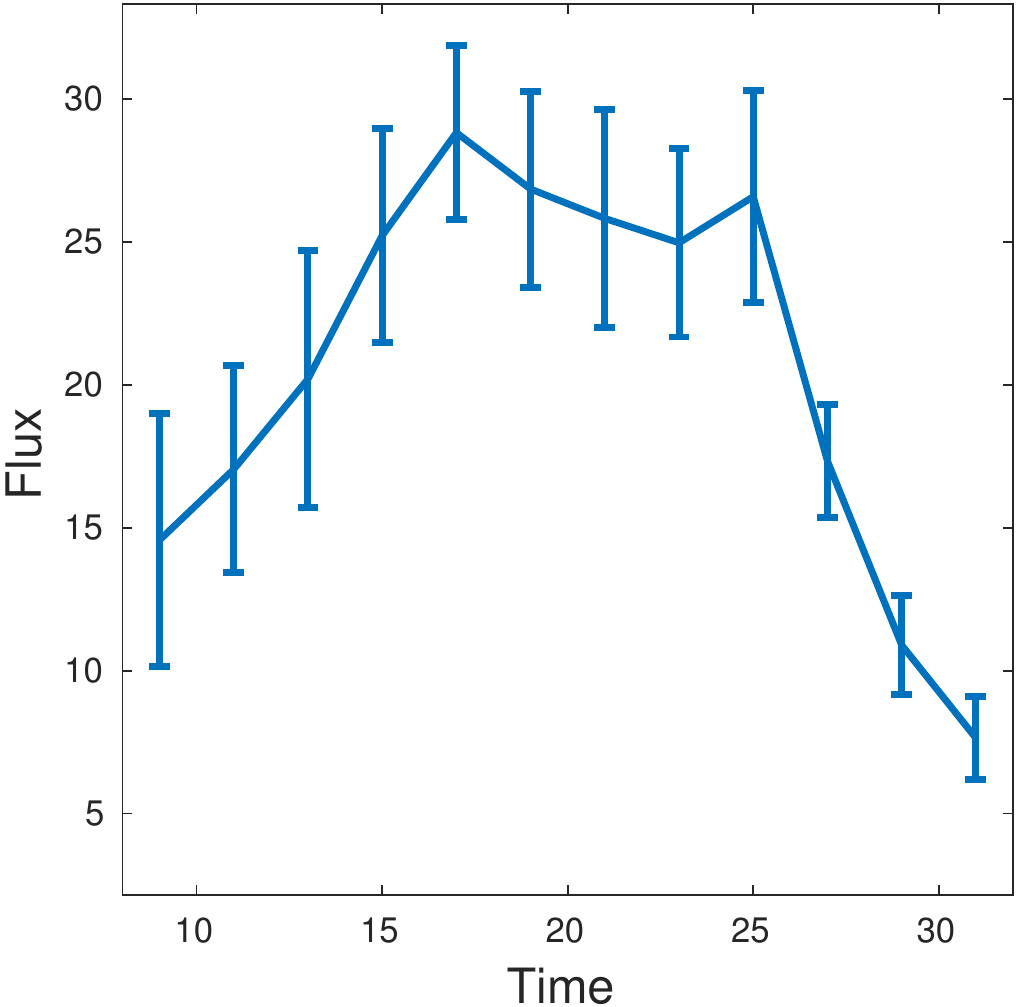}
\end{center}
\caption{Temporal evolution of the $x$ coordinate, the $y$ coordinate and the flux of the southern source. Error bars represent one standard deviation of the posterior distribution. The $x$-axis represents the seconds after 11:06 UT.}
\label{fig:dynamic_param}
\end{figure}

To explore this more fully, it is desirable to examine the source structure at finer time resolution. The shortest practical time resolution for visibility-based imaging using {\em RHESSI} is $\sim$2~s, corresponding to half the satellite rotation period and so the shortest time interval for which a complete set of visibilities can be found.  Figure~3 of \cite{2002SoPh..210..229K} shows results every 2~s, but it should be noted that in the interests of improving the statistics, each image corresponds to a (full rotation) 4~s integration time, so that there is some overlap among the images.  To further explore this evolving structure with our Bayesian method, we computed the source structure with the same 2~s time resolution and 4~s integration time.  Figure~\ref{fig:dynamic} shows the distributions of the particles at the last iteration, and Figure~\ref{fig:dynamic_param} shows the estimated $x$ coordinate (arcseconds West), $y$ coordinate (arcseconds North), and flux (photons s$^{-1}$ cm$^{-2}$) of the Southern footpoint source as functions of time.

The $y$-coordinate of this Southern footpoint, identified as ``Source 2'' by \cite{2002SoPh..210..229K}, has a relatively sudden Southerly (i.e., towards more negative $y$-values) jump around 11:06:13~UT. This is the time at which \cite{2002SoPh..210..229K} suggest that two different Southern footpoint sources may be present.  However, at this time our method (top right panel of Figure~\ref{fig:dynamic}) shows only a small probability of an additional source a few arcseconds to the North of the main source of emission in this region. \cite{2002SoPh..210..229K} identify this weak source as the continuation of ``Source 2''  (which then promptly fades in intensity), and they identify the Southernmost bright patch as the appearance of a new ``Source 3,'' which rises in intensity very quickly (their Figure~4). However, the weak source identified as the continuation of ``Source 2'' by \cite{2002SoPh..210..229K} is very short-lived (only one time interval; Figure~\ref{fig:dynamic}) and therefore most likely a small artifact of noise. We instead contend that the Southernmost source, and not the one just to the North of it, is the continuation of ``Source 2,'' appearing at a slightly new location 3~arcseconds to the South. (This relatively sudden change in location of this source might also be an artifact of noise, considering that we are using a relatively short integration interval.)  The fact that all the parameters ($x$, $y$, and flux) associated with the brightest Southerly source all vary smoothly throughout the event confirms with a high probability that they are all manifestations of a single source in this region. We shall return to the significance of this result in Section~\ref{sec:disc_concl}.

\section{The impact of the prior distribution}
\label{sec:impact}

When using a Bayesian approach to solve an inverse problem, the results may depend on the choice of the prior distribution.
In our tests, we observed that the impact of the prior distribution is negligible whenever the data are highly informative, while, as expected, becomes more relevant when the data carry only little information on the estimated quantity. In order to show this, we present here results obtained by repeated analysis with different values of the prior probabilities for the number of sources and for the source types. 

\subsection{The prior on the number of sources}

In order to assess the sensitivity of the results to the parameter $\lambda$ of the prior, we select two data sets among those already analyzed. For both cases, we let $\lambda$ vary between $0.5$ and $4$ with steps of $0.5$. 

\begin{figure}
	\begin{center}
\subfloat[11:06:15 UT]{\includegraphics[width=8cm]{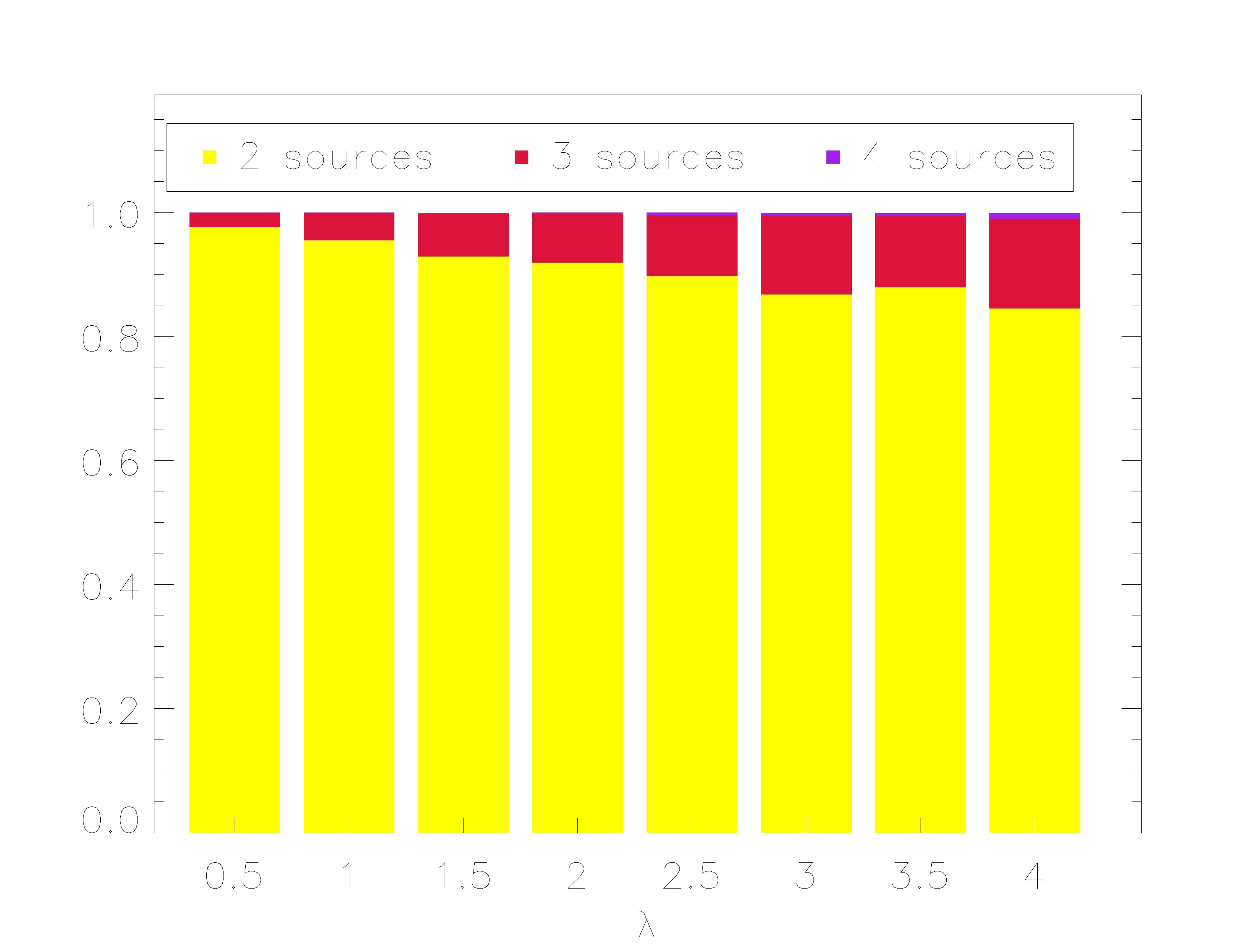} }
\subfloat[20-30 keV]{\includegraphics[width=8cm]{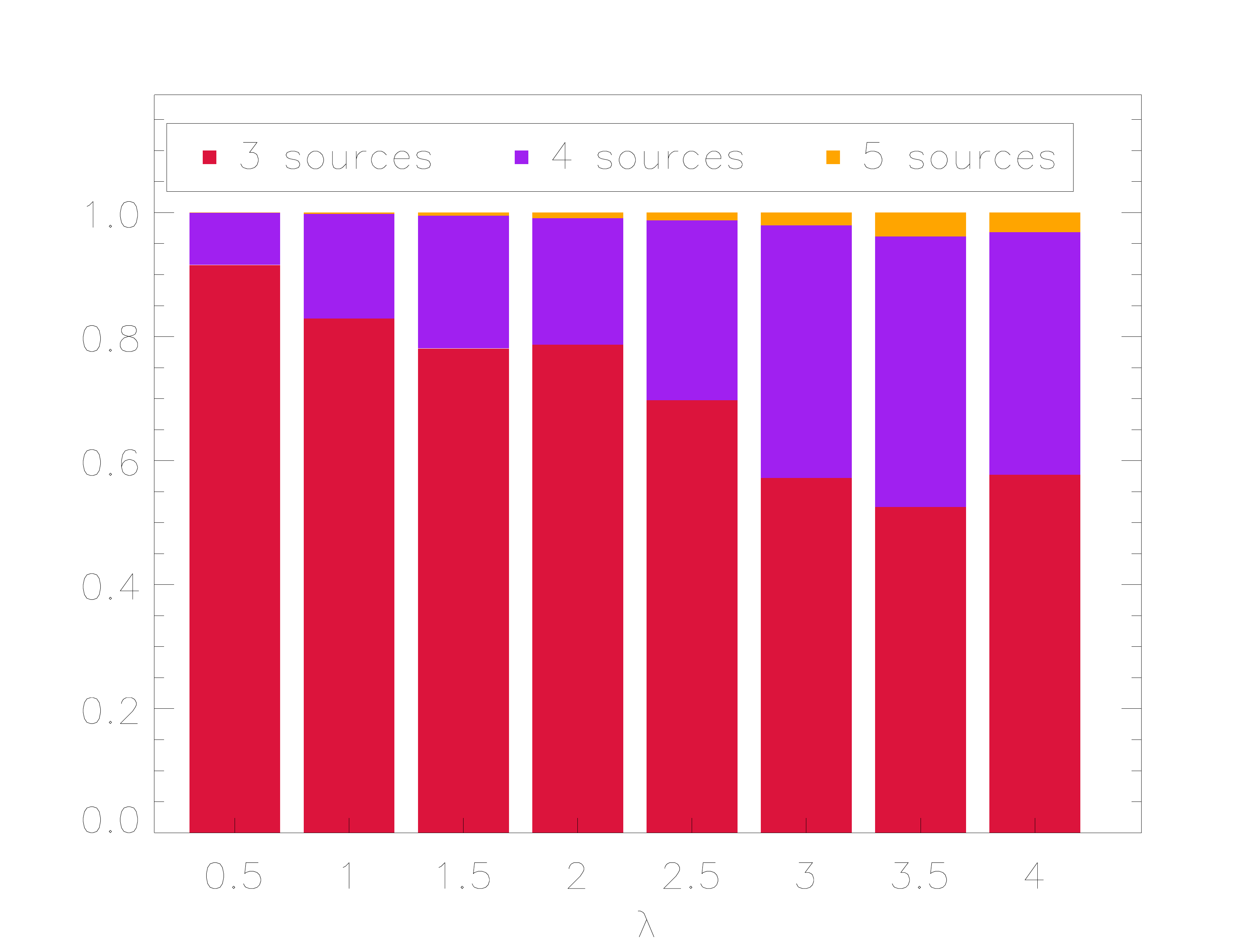}}
	\end{center}
	\caption{Posterior probabilities for the number of sources, for different values of the prior parameter $\lambda$.}
	\label{fig:lambda_krucker}
\end{figure}

The first case was selected as an example of a simple case: it is the 30-80 keV interval, with a 8 second interval centered at 11:06:15 (see Section \ref{sec:kru}). The results are reported in Figure \ref{fig:lambda_krucker}: the posterior probabilities change very little, compared to the prior probabilities, in this case; this is likely due to the good information content of the data.

As an example of a more challenging case, we report the posterior probabilities for the 20-30 keV interval (see Section \ref{sec:sui}). Here the number of sources is higher, and the results are more affected by a change of $\lambda$; specifically, the posterior probability for a four--source configuration reaches $43\%$ when $\lambda = 3.5$ -- still lower than the $53\%$ of the three--source configuration, but close.
On the other hand, the MAP estimate is always a three--source model, even with the lowest $\lambda$.

\subsection{The prior on the source type}

\begin{figure}
	\begin{center}
		\subfloat[$p_C = \frac{1}{2}, \/ p_E = \frac{1}{4}, \/ p_L = \frac{1}{4}$]{\includegraphics[width=7cm]{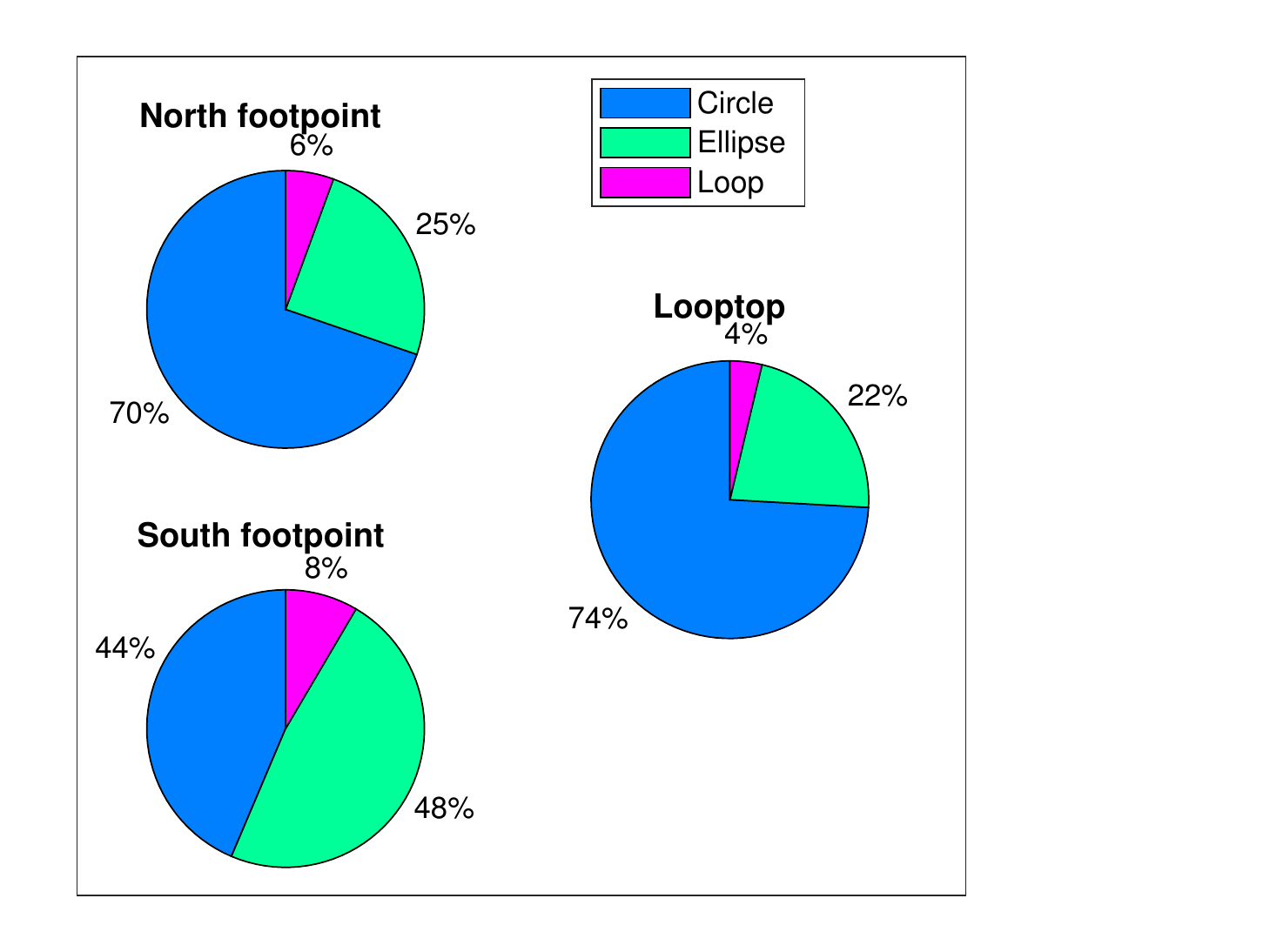}}\hspace{-1.7cm}
		\subfloat[$p_C = \frac{1}{4}, \/ p_E = \frac{1}{2}, \/ p_L = \frac{1}{4}$]{\includegraphics[width=7cm]{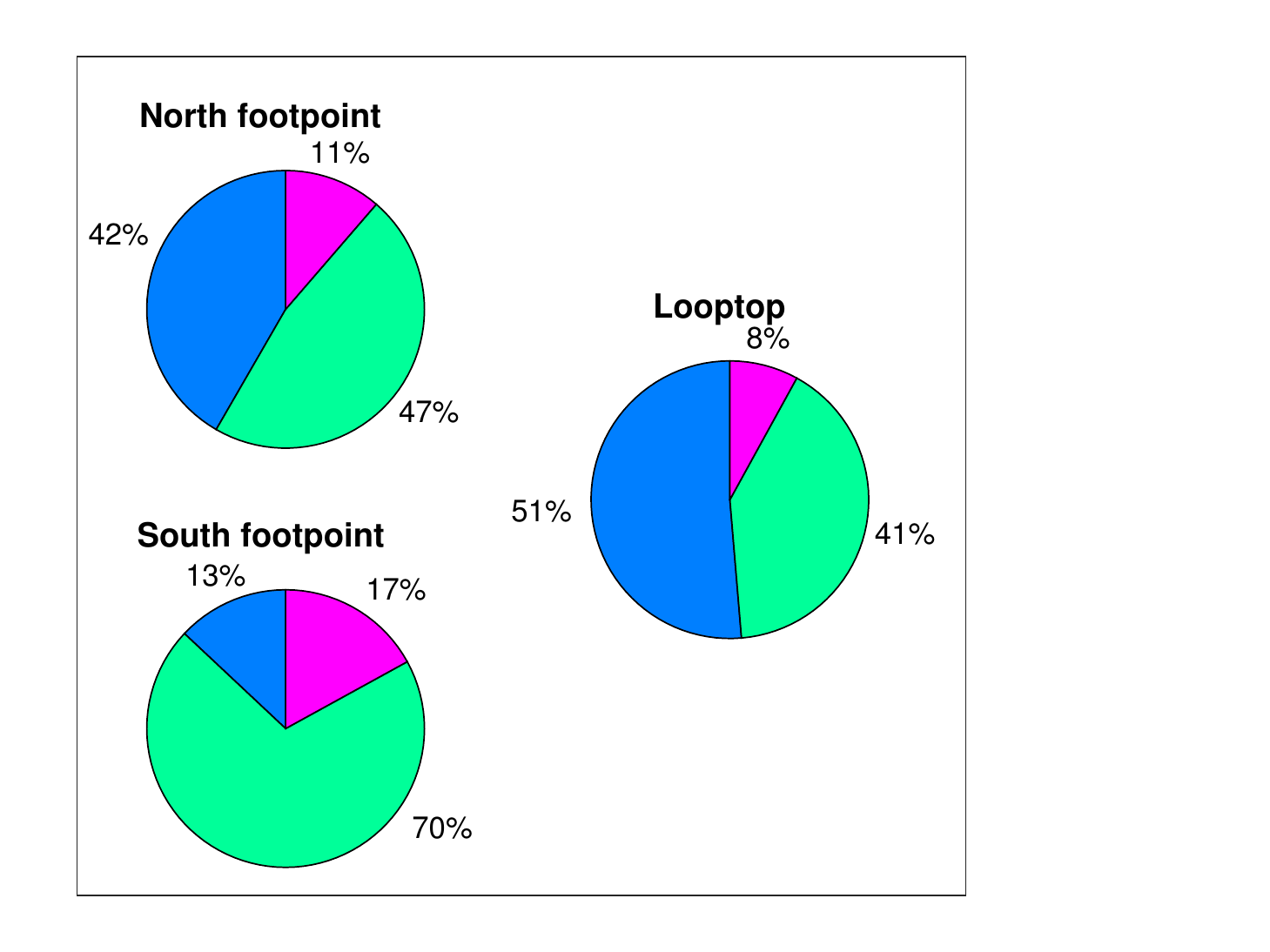}}\hspace{-1.7cm}
		\subfloat[$p_C = \frac{1}{4}, \/ p_E = \frac{1}{4}, \/ p_L = \frac{1}{2}$]{\includegraphics[width=7cm]{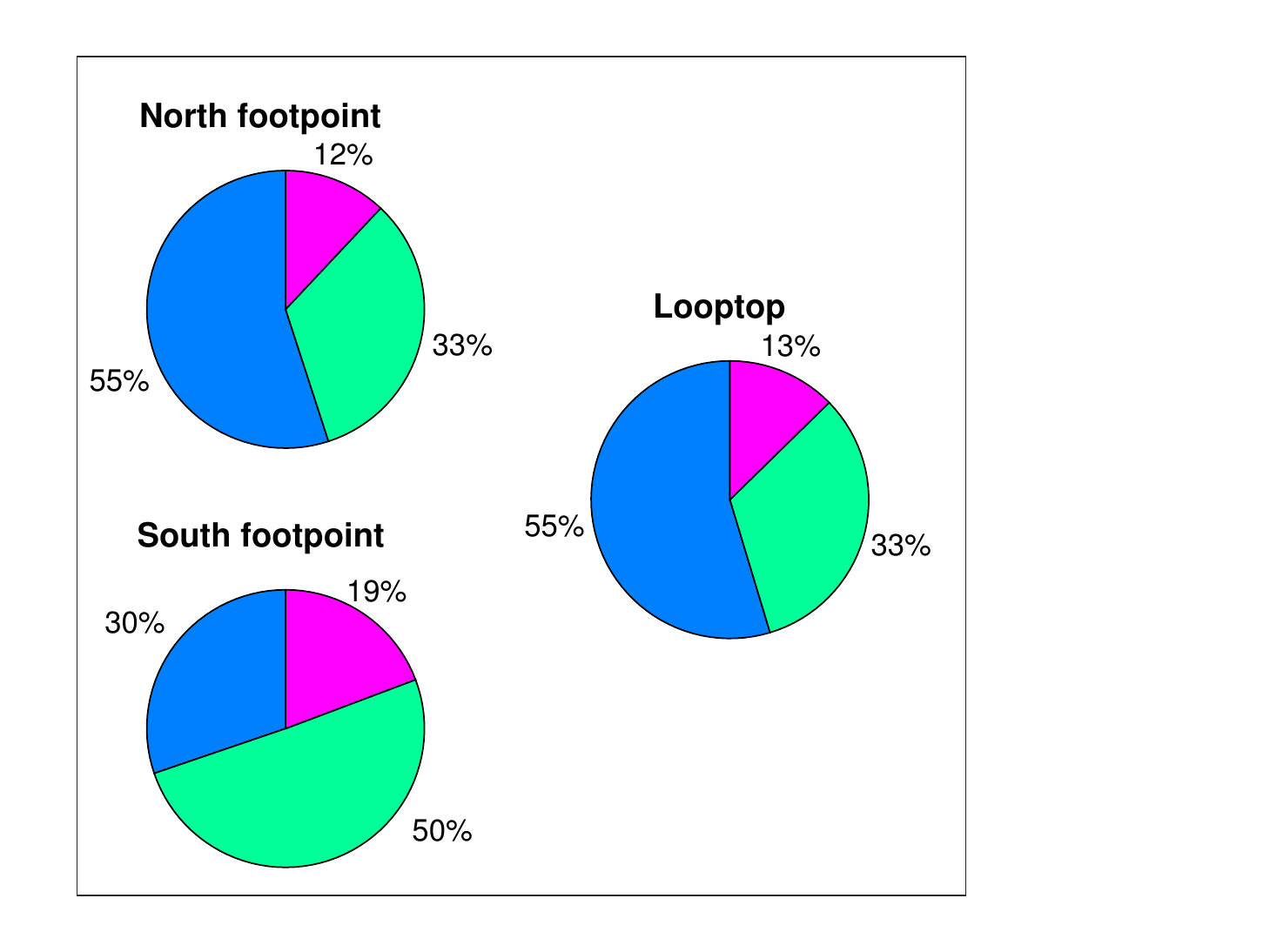}}
	\end{center}
	\caption{Posterior probabilities for the source types, using different a priori probability sets. Left: $p_C = \frac{1}{2}, \/ p_E = \frac{1}{4}, \/ p_L = \frac{1}{4}$. Middle: $p_C = \frac{1}{4}, \/ p_E = \frac{1}{2}, \/ p_L = \frac{1}{4}$. Right: $p_C = \frac{1}{4}, \/ p_E = \frac{1}{4}, \/ p_L = \frac{1}{2}$.}
	\label{fig:types}
\end{figure}

In order to assess the sensitivity of the results to the prior on the source type, we analyzed the 20-30 keV interval (see Section \ref{sec:sui}) using three different sets of values for $p_C, p_E, p_L$ in the prior distribution (see eq. \eqref{eq:pr_sh}). These values were set to favour, each time, a different source type: the preferred source type was given a prior probability of $1/2$, while the other two were given a prior probability of $1/4$. The results are shown in Figure \ref{fig:types}. In each box, we plot the posterior probability for the source type of each estimated source. The results show some expected dependence of the posterior value on the prior, when the source shape is more ambiguous: for instance, the North source is estimated to be a circular source in the first and in the third case, while it is estimated to be an elliptical source in the second case, when the elliptical sources are favoured. On the other hand, the South source is always estimated as an ellipse, though with different probabilities, for all the tested cases. Finally, the loop top is always estimated to be a circular source, though again with different probabilities; however, this result should be taken with care, because the looptop is the weaker source and, as such, the most difficult to estimate; this is also confirmed by the higher uncertainty on its location (cfr. Figure \ref{fig:sui_xy}, panel (d)).

\section{Discussion and conclusions} \label{sec:disc_concl}

The 2002~February~20 event at around 11:06~UT has been the source of intense study by a number of authors  \citep{2002SoPh..210..229K,2002SoPh..210..245S,2002SoPh..210..383A,2002ApJ...580L.177W,2004SoPh..219..149A,2005ApJ...621..482V,2007ApJ...665..846P,2011ApJ...728....4G,2014ApJ...789...71F}. Most of these studies have centered around the morphology of the source, and in particular that of the hard X-ray emission. It is generally agreed that there are several distinct sources of hard X-ray emission in the event, including a loop-top source that dominates at lower energies (presumably indicative of the thermal source), and a number of chromospheric footpoints that dominate at higher energies (presumably associated with the impact of accelerated electrons onto the dense chromosphere).

More interesting is the question as to the number and location of these chromospheric footpoints, in particular whether there are two \citep{2002SoPh..210..245S} or three \citep{2002SoPh..210..229K,2002SoPh..210..261V} such footpoints.  The presence of three separate locations of footpoint emission, especially if each brightens at a different time \citep{2002SoPh..210..229K,2002SoPh..210..261V} would imply a much more complicated magnetic field geometry than a simple two-chromospheric-footpoint-plus-thermal-coronal-source geometry. More specifically, Figure~4 of \cite{2002SoPh..210..229K} (see also Figure~2 of \cite{2002SoPh..210..261V}) suggests a relatively steady Northern footpoint, followed by the appearance of two more southerly footpoints that brighten at different times and appear at different locations.  Although \cite{2002SoPh..210..229K} claim (their Figure~4) that ``Source~3'' (the most Southerly footpoint source) replaces ``Source 2'' (the South footpoint source closest to the Northern footpoint), Figure~4 of that paper \citep[see also][]{2002SoPh..210..261V} shows that ``Source 2'' only gradually diminishes in intensity, and is still present even at the time of peak intensity of ``Source 3.''  It is therefore still a possibility  that ``Source 2'' and ``Source 3'' are simply different manifestations of a single source that moves in a Southerly direction as the event progresses.

So is this event characterized by (1) a Northern footpoint and two separate Southern footpoint sources located at different locations and that brighten at different times (making three footpoint sources in all), or (2) a Northern footpoint and a single Southern source that moves (making two footpoint sources in all)?  The Bayesian method used in this paper is ideally suited to distinguishing between these two possibilities. The results (Sections~3 and~4) show that the most likely scenario is that there are only {\it two} principal sources at any given time, each of which has the highest probability of being a low-eccentricity ($\epsilon \lapprox 0.3$) elliptical source (in one case -- for the Northern source at 11:06:15~UT -- the inferred eccentricity is even sufficiently low that the source is best classified as ``circular.'')

Consistent with the analyses of \cite{2002SoPh..210..229K} and \cite{2002SoPh..210..261V}, we find that the Northern footpoint source is fairly stationary, moving less than 1~arcsecond over the 16~second duration between the first and last images in Figure~5. However, even for the 11:06:15~UT image (middle panel in top row of Figure~5) that is the best case for showing three footpoint sources visible at the same time \citep[Figure~4 of][]{2002SoPh..210..229K}, the Bayesian method reveals that the most likely geometry involves only {\it two} sources, with a single elliptical source (rather than two distinct footpoints) at the Southern location.  Analysis at higher (2~s) time resolution is fully consistent with this interpretation (Figures~\ref{fig:dynamic} and~\ref{fig:dynamic_param}).  While \cite{2002SoPh..210..229K} suggest that two different Southern foopoint sources (``Source 2'' and ``Source 3'') may be present near the peak of the event, the time ($\sim$~11:06:13~UT) at which this interpretation is most convincing corresponds to a time at which the Southern source is split into two components.  One of these, previously identified by \cite{2002SoPh..210..229K} as the continuation of ``Source 2,'' is, however, very weak and short-lived; the other, identified by \cite{2002SoPh..210..229K} as the sudden appearance of a new ``Source 3,'' has a location, size, and intensity that, given the statistical uncertainties, is more probably consistent with a continuation of ``Source 2'' (Figure~\ref{fig:dynamic_param}).  Our analysis therefore favors an interpretation in which the weak source is to be disregarded, so that the emission in the Southern part of the flare arises from only one footpoint, with smoothly varying size, intensity, and location.

This Southern source does move quite rapidly (left and middle panels of Figure~\ref{fig:dynamic_param}), drifting about 7~arc seconds in a South-Southwest direction over a period of about 15~seconds, corresponding to a velocity $\sim$350~km~s$^{-1}$. \cite{2003ApJ...595L.103K} studied footpoint motions in the 2003~July~23 ($\sim$00:30~UT) event; Figure~3 of that paper shows that the fastest motions (in the Northern footpoint of that event) had a speed of order 50~km~s$^{-1}$ \citep[comparable to those inferred by][]{1998ASSL..229..273S}.  The relative motion of the various footpoints in the 2002~July~23 event is qualitatively similar to what we have found for the 2002~February~20 event; the motion in both events is such that the inter-footpoint separation increases with time.  Such a behavior is consistent with the successive activation of higher and higher field lines in an arcade, as in the well-known CSHKP model \citep{1964NASSP..50..451C,1966Natur.211..695S,1974SoPh...34..323H,1976SoPh...50...85K}.

\cite{2003ApJ...595L.103K} also showed that the speed of the footpoint motion in the 2003~July~23 event correlates well with the hard X-ray intensity of the footpoint source, suggesting that the rate of electron acceleration is correlated with the speed of the footpoint motion. Indeed higher footpoint speeds result in higher induced electric fields perpendicular to both the magnetic field direction and the direction of motion: $\mathbf {E} = (\mathbf{v}/c) \times \mathbf{B}$.  Since the 2002~February~20 event was fairly small (a GOES C-class) flare, we take B = 100~G as a representative value; taking $V = 10^{27}$~cm$^{-3}$, the associated energy content $(B^2/8\pi) V$ in the magnetic field is then of order $10^{29}$~ergs, consistent with the GOES classification.

The electric field associated with the inferred 350~km~s$^{-1}$ velocity is thus $E \simeq 0.12$~statvolt~cm$^{-1}$ $\simeq 35$~V~cm$^{-1}$.  This is several orders of magnitude larger than the \cite{1959PhRv..115..238D} field $E_D \simeq 10^{-8} n/T \simeq 10^{-5}$~V~cm$^{-1}$ (using $n \simeq 10^{10}$~cm$^{-3}$ and $T \simeq 10^7$~K) that is necessary to accelerate electrons out of a background plasma. It is also sufficiently large that it can accelerate electrons to the required $\simeq$30~keV energies in a distance less than 10 meters, which is orders of magnitude smaller than the characteristic size of a flare loop.  Such electric field magnitudes and acceleration lengths are, however, nicely compatible with those associated with super-Dreicer acceleration in current sheets \citep[see, e.g.,][]{1992ASSL..172.....S,1996ApJ...462..997L}.


\section*{Acknowledgements}
AGE was supported by grant NNX17AI16G from NASA's Heliophysics Supporting Research program. FS, AS, AMM and MP have been supported by the H2020 grant Flare Likelihood And Region Eruption foreCASTing (FLARECAST), project number 640216. The authors kindly thank Dr. Kim Tolbert for providing access to results from previous studies.
\section*{Appendix: The Adaptive Sequential Monte Carlo algorithm}

The main reason to introduce SMC samplers is that, due to the complexity of the posterior distribution, drawing random samples from it is not trivial, so that basic Monte Carlo approaches would either fail or be extremely inefficient.
The main idea behind SMC samplers is to construct an auxiliary sequence of distributions $\{\pi_i(x)\}_{i=1,\dots,I}$, such that the first distribution is the prior distribution $\pi_1(X)=\pi_{\text{pr}}(X)$, the last distribution is the target posterior distribution $\pi_I(X)=\pi_{\text{po}}(X|y)$, and the sequence is ``smooth.''  We draw a random sample from $\pi_1(x)$, make the corresponding sample points evolve according to properly selected Markov kernels, and re-weigh them; eventually, the sample points at the final iteration $I$ will be distributed according to the target distribution, which in our case is the posterior distribution.

At the $i$-th iteration we use

\begin{equation}
\label{eq:po}
\pi_i(X)= \mathcal{L}(y|X)^{\gamma_i}\pi_{\text{pr}}(X),
\end{equation}
where $\gamma_1=0 $, $\gamma_{I}=1$ and $\gamma_i \in [0,1]$, with $\gamma_{i+1} > \gamma_i, \forall i$. In other words, we start from the prior distribution and we go progressively towards the posterior distribution by systematically increasing the exponent $\gamma_i$.

The choice of the $\gamma_i$ is very important for the success of the algorithm; here we employ an adaptive approach in which the sequence of the distributions is determined on-line, with an empirical measure of the distance between two subsequent distributions (i.e., it depends on how much the distribution at the iteration $(i+1)$ differs from the one at the iteration $i$). For more details we refer the reader to \cite{sorrentino2014bayesian}.

Since the Bayesian model considered here has a variable dimension (see Equation~\eqref{eq:space}), we need to introduce the concept of {\it reversible jump moves}, which allow us to explore the space $\chi$ of the parameters with which flares are modeled. In our analysis we consider the following moves:

\begin{itemize}
\item birth and death;
\item change;
\item split and merge; and
\item update.
\end{itemize}
Birth and death moves are attempted with probability $1/3$. The type of the source is chosen according to $\rho_2$ in Equation~\eqref{eq:pr_sh} and the parameters of the newly born source are uniformly distributed following Equation~\eqref{eq:pr_par}. A change move is done during each iteration and transforms a circle into an ellipse (and vice versa), or an ellipse into a loop (and vice versa). (We do not consider the transition from a circle to a loop (and vice versa) since we want to have smooth transformations.) Split and merge moves are applied only to the ellipses and allow one ellipse to split in two (or two ellipses to merge in one) and are attempted at each iteration. The update move is a classical move of the Metropolis-Hastings algorithm and it is done at the end of each iteration. This move does not change the type of the source, but it does determine how each parameter of each source evolves independently from the others

In summary, this process allows the description of physical situations in which flaring sources may appear, disappear, and change their shape and physical properties. From the mathematical viewpoint, the split, merge and change moves are introduced here for the first time.  

In Table \ref{alg:ASCMC}, we give a sketch of the ASMC algorithm. For more details, we refer the reader to  \cite{del2006sequential}.


\begin{table}
\caption{ASMC algorithm}\label{alg:ASCMC}
\begin{algorithmic}
\STATE{\textit{Initialization of the sample}}
\FOR{$p=1,\dots, N$}
\STATE{draw $X_0^{p}$ from $\pi_{\text{pr}}$}
\ENDFOR
\STATE{Set $i=1$ and $\gamma_1=0$}
\WHILE{$\gamma_i \leq 1$}
\STATE{Apply a possible resampling}
\FOR{each particle } 
\STATE{propose birth/death/split/merge/change moves, then accept/reject}
\STATE{for each parameter, update the value, then accept/reject}
\STATE{compute the weights}
\ENDFOR
\STATE{Adaptive determination of the next exponent $\gamma_{i+1}$}
\STATE{$i \to i+1$}
\ENDWHILE
\RETURN $X$
\end{algorithmic}
\end{table}

\bibliographystyle{aasjournal}

\bibliography{references}

\begin{thebibliography}{}
\expandafter\ifx\csname natexlab\endcsname\relax\def\natexlab#1{#1}\fi
\providecommand{\url}[1]{\href{#1}{#1}}

\bibitem[{{Aschwanden} {et~al.}(2002{\natexlab{a}}){Aschwanden}, {Brown}, \&
  {Kontar}}]{2002SoPh..210..383A}
{Aschwanden}, M.~J., {Brown}, J.~C., \& {Kontar}, E.~P. 2002{\natexlab{a}},
  \solphys, 210, 383

\bibitem[{{Aschwanden} {et~al.}(2004){Aschwanden}, {Metcalf}, {Krucker},
  {Sato}, {Conway}, {Hurford}, \& {Schmahl}}]{2004SoPh..219..149A}
{Aschwanden}, M.~J., {Metcalf}, T.~R., {Krucker}, S., {et~al.} 2004, \solphys,
  219, 149

\bibitem[{{Aschwanden} {et~al.}(2002{\natexlab{b}}){Aschwanden}, {Schmahl}, \&
  {RHESSI Team}}]{2002SoPh..210..193A}
{Aschwanden}, M.~J., {Schmahl}, E., \& {RHESSI Team}. 2002{\natexlab{b}},
  \solphys, 210, 193

\bibitem[{{Benvenuto} {et~al.}(2013){Benvenuto}, {Schwartz}, {Piana}, \&
  {Massone}}]{2013A&A...555A..61B}
{Benvenuto}, F., {Schwartz}, R., {Piana}, M., \& {Massone}, A.~M. 2013, \aap,
  555, A61

\bibitem[{{Bong} {et~al.}(2006){Bong}, {Lee}, {Gary}, \&
  {Yun}}]{2006ApJ...636.1159B}
{Bong}, S.-C., {Lee}, J., {Gary}, D.~E., \& {Yun}, H.~S. 2006, \apj, 636, 1159

\bibitem[{{Brown}(1971)}]{1971SoPh...18..489B}
{Brown}, J.~C. 1971, \solphys, 18, 489

\bibitem[{{Brown}(1972)}]{1972SoPh...26..441B}
---. 1972, \solphys, 26, 441

\bibitem[{{Brown}(1973)}]{1973SoPh...31..143B}
---. 1973, \solphys, 31, 143

\bibitem[{{Brown} {et~al.}(2002){Brown}, {Aschwanden}, \&
  {Kontar}}]{2002SoPh..210..373B}
{Brown}, J.~C., {Aschwanden}, M.~J., \& {Kontar}, E.~P. 2002, \solphys, 210,
  373

\bibitem[{{Brown} \& {McClymont}(1975)}]{1975SoPh...41..135B}
{Brown}, J.~C., \& {McClymont}, A.~N. 1975, \solphys, 41, 135

\bibitem[{{Brown} {et~al.}(2009){Brown}, {Turkmani}, {Kontar}, {MacKinnon}, \&
  {Vlahos}}]{2009A&A...508..993B}
{Brown}, J.~C., {Turkmani}, R., {Kontar}, E.~P., {MacKinnon}, A.~L., \&
  {Vlahos}, L. 2009, \aap, 508, 993

\bibitem[{{Carmichael}(1964)}]{1964NASSP..50..451C}
{Carmichael}, H. 1964, NASA Special Publication, 50, 451

\bibitem[{Del~Moral {et~al.}(2006)Del~Moral, Doucet, \&
  Jasra}]{del2006sequential}
Del~Moral, P., Doucet, A., \& Jasra, A. 2006, Journal of the Royal Statistical
  Society: Series B (Statistical Methodology), 68, 411

\bibitem[{{Delaboudini{\`e}re} {et~al.}(1995){Delaboudini{\`e}re}, {Artzner},
  {Brunaud}, {Gabriel}, {Hochedez}, {Millier}, {Song}, {Au}, {Dere}, {Howard},
  {Kreplin}, {Michels}, {Moses}, {Defise}, {Jamar}, {Rochus}, {Chauvineau},
  {Marioge}, {Catura}, {Lemen}, {Shing}, {Stern}, {Gurman}, {Neupert},
  {Maucherat}, {Clette}, {Cugnon}, \& {van Dessel}}]{1995SoPh..162..291D}
{Delaboudini{\`e}re}, J.-P., {Artzner}, G.~E., {Brunaud}, J., {et~al.} 1995,
  \solphys, 162, 291

\bibitem[{{Dennis} \& {Pernak}(2009)}]{2009ApJ...698.2131D}
{Dennis}, B.~R., \& {Pernak}, R.~L. 2009, \apj, 698, 2131

\bibitem[{{Dreicer}(1959)}]{1959PhRv..115..238D}
{Dreicer}, H. 1959, Physical Review, 115, 238

\bibitem[{Duval-Poo {et~al.}(2017)Duval-Poo, Massone, \&
  Piana}]{duval2017compressed}
Duval-Poo, M.~A., Massone, A.~M., \& Piana, M. 2017, in Sampling Theory and
  Applications (SampTA), 2017 International Conference on, IEEE, 677--681

\bibitem[{{Duval-Poo} {et~al.}(2017){Duval-Poo}, {Piana}, \&
  {Massone}}]{2017arXiv170803877D}
{Duval-Poo}, M.~A., {Piana}, M., \& {Massone}, A.~M. 2017, ArXiv e-prints,
  arXiv:1708.03877

\bibitem[{{Emslie}(1978)}]{1978ApJ...224..241E}
{Emslie}, A.~G. 1978, \apj, 224, 241

\bibitem[{{Emslie} {et~al.}(2003){Emslie}, {Kontar}, {Krucker}, \&
  {Lin}}]{2003ApJ...595L.107E}
{Emslie}, A.~G., {Kontar}, E.~P., {Krucker}, S., \& {Lin}, R.~P. 2003, \apjl,
  595, L107

\bibitem[{{Emslie} \& {Machado}(1987)}]{1987SoPh..107..263E}
{Emslie}, A.~G., \& {Machado}, M.~E. 1987, \solphys, 107, 263

\bibitem[{{Falewicz}(2014)}]{2014ApJ...789...71F}
{Falewicz}, R. 2014, \apj, 789, 71

\bibitem[{{Felix} {et~al.}(2017){Felix}, {Bolzern}, \&
  {Battaglia}}]{2017ApJ...849...10F}
{Felix}, S., {Bolzern}, R., \& {Battaglia}, M. 2017, \apj, 849, 10

\bibitem[{{Guo} {et~al.}(2012{\natexlab{a}}){Guo}, {Emslie}, {Kontar},
  {Benvenuto}, {Massone}, \& {Piana}}]{2012A&A...543A..53G}
{Guo}, J., {Emslie}, A.~G., {Kontar}, E.~P., {et~al.} 2012{\natexlab{a}}, \aap,
  543, A53

\bibitem[{{Guo} {et~al.}(2012{\natexlab{b}}){Guo}, {Emslie}, {Massone}, \&
  {Piana}}]{2012ApJ...755...32G}
{Guo}, J., {Emslie}, A.~G., {Massone}, A.~M., \& {Piana}, M.
  2012{\natexlab{b}}, \apj, 755, 32

\bibitem[{{Guo} {et~al.}(2011){Guo}, {Liu}, {Fletcher}, \&
  {Kontar}}]{2011ApJ...728....4G}
{Guo}, J., {Liu}, S., {Fletcher}, L., \& {Kontar}, E.~P. 2011, \apj, 728, 4

\bibitem[{{Hirayama}(1974)}]{1974SoPh...34..323H}
{Hirayama}, T. 1974, \solphys, 34, 323

\bibitem[{{H{\"o}gbom}(1974)}]{1974A&AS...15..417H}
{H{\"o}gbom}, J.~A. 1974, \aaps, 15, 417

\bibitem[{{Hurford} {et~al.}(2002){Hurford}, {Schmahl}, {Schwartz}, {Conway},
  {Aschwanden}, {Csillaghy}, {Dennis}, {Johns-Krull}, {Krucker}, {Lin},
  {McTiernan}, {Metcalf}, {Sato}, \& {Smith}}]{2002SoPh..210...61H}
{Hurford}, G.~J., {Schmahl}, E.~J., {Schwartz}, R.~A., {et~al.} 2002, \solphys,
  210, 61

\bibitem[{{Kontar} {et~al.}(2008){Kontar}, {Hannah}, \&
  {MacKinnon}}]{2008A&A...489L..57K}
{Kontar}, E.~P., {Hannah}, I.~G., \& {MacKinnon}, A.~L. 2008, \aap, 489, L57

\bibitem[{{Kopp} \& {Pneuman}(1976)}]{1976SoPh...50...85K}
{Kopp}, R.~A., \& {Pneuman}, G.~W. 1976, \solphys, 50, 85

\bibitem[{{Krucker} {et~al.}(2003){Krucker}, {Hurford}, \&
  {Lin}}]{2003ApJ...595L.103K}
{Krucker}, S., {Hurford}, G.~J., \& {Lin}, R.~P. 2003, \apjl, 595, L103

\bibitem[{{Krucker} \& {Lin}(2002)}]{2002SoPh..210..229K}
{Krucker}, S., \& {Lin}, R.~P. 2002, \solphys, 210, 229

\bibitem[{{Krucker} \& {Lin}(2008)}]{2008ApJ...673.1181K}
---. 2008, \apj, 673, 1181

\bibitem[{{Krucker} {et~al.}(2007){Krucker}, {White}, \&
  {Lin}}]{2007ApJ...669L..49K}
{Krucker}, S., {White}, S.~M., \& {Lin}, R.~P. 2007, \apjl, 669, L49

\bibitem[{{Krucker} {et~al.}(2008){Krucker}, {Battaglia}, {Cargill},
  {Fletcher}, {Hudson}, {MacKinnon}, {Masuda}, {Sui}, {Tomczak}, {Veronig},
  {Vlahos}, \& {White}}]{2008A&ARv..16..155K}
{Krucker}, S., {Battaglia}, M., {Cargill}, P.~J., {et~al.} 2008, \aapr, 16, 155

\bibitem[{{Lin} {et~al.}(2002){Lin}, {Dennis}, {Hurford}, {Smith}, {Zehnder},
  {Harvey}, {Curtis}, {Pankow}, {Turin}, {Bester}, {Csillaghy}, {Lewis},
  {Madden}, {van Beek}, {Appleby}, {Raudorf}, {McTiernan}, {Ramaty}, {Schmahl},
  {Schwartz}, {Krucker}, {Abiad}, {Quinn}, {Berg}, {Hashii}, {Sterling},
  {Jackson}, {Pratt}, {Campbell}, {Malone}, {Landis}, {Barrington-Leigh},
  {Slassi-Sennou}, {Cork}, {Clark}, {Amato}, {Orwig}, {Boyle}, {Banks},
  {Shirey}, {Tolbert}, {Zarro}, {Snow}, {Thomsen}, {Henneck}, {McHedlishvili},
  {Ming}, {Fivian}, {Jordan}, {Wanner}, {Crubb}, {Preble}, {Matranga}, {Benz},
  {Hudson}, {Canfield}, {Holman}, {Crannell}, {Kosugi}, {Emslie}, {Vilmer},
  {Brown}, {Johns-Krull}, {Aschwanden}, {Metcalf}, \&
  {Conway}}]{2002SoPh..210....3L}
{Lin}, R.~P., {Dennis}, B.~R., {Hurford}, G.~J., {et~al.} 2002, \solphys, 210,
  3

\bibitem[{{Litvinenko}(1996)}]{1996ApJ...462..997L}
{Litvinenko}, Y.~E. 1996, \apj, 462, 997

\bibitem[{{Massone} {et~al.}(2009){Massone}, {Emslie}, {Hurford}, {Prato},
  {Kontar}, \& {Piana}}]{2009ApJ...703.2004M}
{Massone}, A.~M., {Emslie}, A.~G., {Hurford}, G.~J., {et~al.} 2009, \apj, 703,
  2004

\bibitem[{{Masuda} {et~al.}(1995){Masuda}, {Kosugi}, {Hara}, {Sakao},
  {Shibata}, \& {Tsuneta}}]{1995PASJ...47..677M}
{Masuda}, S., {Kosugi}, T., {Hara}, H., {et~al.} 1995, \pasj, 47, 677

\bibitem[{{Metcalf} {et~al.}(1996){Metcalf}, {Hudson}, {Kosugi}, {Puetter}, \&
  {Pina}}]{1996ApJ...466..585M}
{Metcalf}, T.~R., {Hudson}, H.~S., {Kosugi}, T., {Puetter}, R.~C., \& {Pina},
  R.~K. 1996, \apj, 466, 585

\bibitem[{{Piana} {et~al.}(2007){Piana}, {Massone}, {Hurford}, {Prato},
  {Emslie}, {Kontar}, \& {Schwartz}}]{2007ApJ...665..846P}
{Piana}, M., {Massone}, A.~M., {Hurford}, G.~J., {et~al.} 2007, \apj, 665, 846

\bibitem[{{Sakao} {et~al.}(1998){Sakao}, {Kosugi}, \&
  {Masuda}}]{1998ASSL..229..273S}
{Sakao}, T., {Kosugi}, T., \& {Masuda}, S. 1998, in Astrophysics and Space
  Science Library, Vol. 229, Observational Plasma Astrophysics : Five Years of
  YOHKOH and Beyond, ed. T.~{Watanabe} \& T.~{Kosugi}, 273

\bibitem[{{Sato}(2004)}]{2004cosp...35.3787S}
{Sato}, J. 2004, in COSPAR Meeting, Vol.~35, 35th COSPAR Scientific Assembly,
  ed. J.-P. {Paill{\'e}}, 3787

\bibitem[{{Sato} {et~al.}(1999){Sato}, {Kosugi}, \&
  {Makishima}}]{1999PASJ...51..127S}
{Sato}, J., {Kosugi}, T., \& {Makishima}, K. 1999, \pasj, 51, 127

\bibitem[{{Somov}(1992)}]{1992ASSL..172.....S}
{Somov}, B.~V., ed. 1992, Astrophysics and Space Science Library, Vol. 172,
  {Physical processes in solar flares.}

\bibitem[{Sorrentino {et~al.}(2014)Sorrentino, Luria, \&
  Aramini}]{sorrentino2014bayesian}
Sorrentino, A., Luria, G., \& Aramini, R. 2014, Inverse Problems, 30, 045010

\bibitem[{{Sturrock}(1966)}]{1966Natur.211..695S}
{Sturrock}, P.~A. 1966, \nat, 211, 695

\bibitem[{{Sui} {et~al.}(2002){Sui}, {Holman}, {Dennis}, {Krucker}, {Schwartz},
  \& {Tolbert}}]{2002SoPh..210..245S}
{Sui}, L., {Holman}, G.~D., {Dennis}, B.~R., {et~al.} 2002, \solphys, 210, 245

\bibitem[{{Veronig} {et~al.}(2005){Veronig}, {Brown}, {Dennis}, {Schwartz},
  {Sui}, \& {Tolbert}}]{2005ApJ...621..482V}
{Veronig}, A.~M., {Brown}, J.~C., {Dennis}, B.~R., {et~al.} 2005, \apj, 621,
  482

\bibitem[{{Vilmer} {et~al.}(2002){Vilmer}, {Krucker}, {Lin}, \& {Rhessi
  Team}}]{2002SoPh..210..261V}
{Vilmer}, N., {Krucker}, S., {Lin}, R.~P., \& {Rhessi Team}. 2002, \solphys,
  210, 261

\bibitem[{{Wang} {et~al.}(2002){Wang}, {Ji}, {Schmahl}, {Qiu}, {Liu}, \&
  {Deng}}]{2002ApJ...580L.177W}
{Wang}, H., {Ji}, H., {Schmahl}, E.~J., {et~al.} 2002, \apjl, 580, L177

\bibitem[{{Xu} {et~al.}(2008){Xu}, {Emslie}, \&
  {Hurford}}]{2008ApJ...673..576X}
{Xu}, Y., {Emslie}, A.~G., \& {Hurford}, G.~J. 2008, \apj, 673, 576

\end{thebibliography}

\end{document}